\documentclass[journal]{IEEEtran}
\usepackage{amsmath,amsfonts}
\usepackage{algorithmic}
\usepackage{array}
\usepackage[caption=false,font=normalsize,labelfont=sf,textfont=sf]{subfig}
\usepackage{textcomp}
\usepackage{stfloats}
\usepackage{url}
\usepackage{verbatim}
\usepackage{graphicx}
\usepackage{enumitem}
\usepackage{multirow}
\usepackage{cite}
\def\BibTeX{{\rm B\kern-.05em{\sc i\kern-.025em b}\kern-.08em
    T\kern-.1667em\lower.7ex\hbox{E}\kern-.125emX}}
\usepackage{balance}
\usepackage[utf8]{inputenc}
\usepackage{newunicodechar}
\newunicodechar{−}{\textminus}
\begin{document}

\title{What You Train Is What You Get: Gender Bias, Training Composition, and Post-Hoc Mitigation in Audio Deepfake Detection}

\author{Aishwarya~R.~Fursule,~\IEEEmembership{}
        Vamshi~Nallaguntla,~\IEEEmembership{}
        Shruti~Kshirsagar,~\IEEEmembership{}
        and~Anderson~R.~Avila~\IEEEmembership{}

\thanks{A. Fursule, V. Nallaguntla, and S. Kshirsagar are with the School of Computing, 
Wichita State University, Wichita, KS, USA 
(e-mail: axfursule@shockers.wichita.edu; vxnallaguntla@shockers.wichita.edu; shruti.kshirsagar@wichita.edu).}
\thanks{A. R. Avila is with the Institut national de la recherche 
scientifique (INRS--EMT), Montreal, QC, Canada, and also with the 
INRS-UQO Mixed Research Unit on Cybersecurity, Gatineau, QC, Canada 
(e-mail: anderson.avila@inrs.ca).}}

\maketitle
\begin{abstract}
Audio deepfake detection models determine whether speech is genuine or artificially generated, but high overall accuracy can mask substantial performance disparities across demographic groups. In this work, we investigate gender bias in audio deepfake detection using the ASVspoof5 dataset. We use ASVspoof5 under a controlled custom split designed to isolate gender-composition effects. We train attack-specific models on nine training sets with different gender compositions, ranging from female-only to male-only. We use a ResNet18 classifier with LogSpectrogram and WavLM-Base+ features, and we evaluated six post-hoc threshold calibration methods. Experimental results show that training data 
composition strongly predicts bias direction, with the 
underrepresented gender performing worse at test time. WavLM-Base+ features are shown to 
produce gender performance gaps 3.0 to 4.3 times larger 
than LogSpectrogram under identical training conditions, 
and balanced training is found to reduce LogSpectrogram 
bias but leave WavLM bias largely intact. Moreover, all 
six calibration strategies, including Oracle calibration 
with full test-set label access, leave the Equal Error 
Rate gap unchanged at 1.317\,pp, confirming that 
threshold adjustment cannot correct underlying score 
distribution disparities. Overall, these findings 
suggest that gender fairness in audio deepfake detection 
must be addressed at training time, as post-hoc methods 
can only partially mitigate the resulting disparities.
\end{abstract}

\begin{IEEEkeywords}
Audio deepfake detection, gender bias, training data composition, threshold calibration, spoofing attacks, trustworthy AI
\end{IEEEkeywords}

\section{Introduction}

\IEEEPARstart{S}{peech} can no longer be reliably trusted as an indicator of human identity \cite{mai2023humans}. Advances in neural text-to-speech synthesis, voice conversion, and adversarial perturbation now enable generation of audio that can convincingly mimic human speech and deceive listeners~\cite{c1, c2}. The research community has responded with a range of countermeasures, from classical Gaussian Mixture Model classifiers to end-to-end architectures such as RawNet2~\cite{c6}, graph attention networks such as AASIST~\cite{c7}, and self-supervised representations including WavLM~\cite{chen2022wavlm}, all benchmarked through the ASVspoof challenge series~\cite{wu2017asvspoof, yamagishi2021asvspoof, liu2023asvspoof, c4}. However, strong average accuracy does not guarantee equitable performance across demographic groups~\cite{hutiri2022bias} as a model may achieve a low aggregate Equal Error Rate (EER) while consistently failing one gender on specific attack types, implying that some users receive less protection based on gender alone.
 
Prior work has established that this failure mode exists but has not resolved its cause, particularly with respect to how attack type, speaker gender, and feature representation interact \cite{yi2023audio, zhang2025audio, hong2026vulnerabilities}. Detection systems trained on female voices outperform those trained on male voices~\cite{c15}, attributed to spectral artifacts in synthesized female speech.
A fairness audit of six detection systems~\cite{yadav2024fairssd} found the opposite: male speakers face higher false positive rates regardless of architecture. A third study~\cite{c21} attributes disparities to score distribution shift between training and evaluation rather than to spectral artifacts. These findings contradict each other because none controlled the gender composition of the training data. It therefore remains unknown whether bias direction is predictable from composition, whether this relationship holds across attack types, and whether the method of achieving gender balance matters. In~\cite{c18} we showed that a single threshold obscures statistically significant gender differences across six fairness metrics on ASVspoof5. In~\cite{fursule2026smc} we identified score distribution asymmetry and embedding-level gender leakage as structural bias sources, and showed that per-gender threshold calibration reduces false alarm disparity by 54 to 75\% without degrading accuracy. Those reductions were in false positive rate gap under balanced training; the present study shows that the EER gap is insensitive to all post-hoc calibration including Oracle calibration \cite{hardt2016equality, chouldechova2017fair} with full test-set label access. Critically, neither study varied training gender composition, leaving the three questions above unresolved.
 
A related question follows naturally from this gap in the literature: is there a fundamental limit to how much post-hoc calibration alone can reduce gender disparities? While group fairness criteria are known to be mutually incompatible~\cite{hardt2016equality,chouldechova2017fair}, whether this extends to a hard post-hoc boundary on the EER gap across diverse training compositions has not been shown. Here, we address these questions by training 288 WavLM-Base+ \cite{chen2022wavlm} models on ASVspoof5 under nine gender-composition configurations, together with 96 LogSpectrogram models under three baseline configurations, for 384 attack-specific models in total. WavLM-Base+ was selected because it exhibited substantially larger gender gaps than other architectures in our earlier work \cite{fursule2026smc}, making it the more informative choice for studying composition effects at scale. Each model is evaluated separately on female and male speakers across all 32 spoofing attacks, with six group fairness metrics assessed under six post-hoc calibration strategies, including an Oracle calibration with direct access to evaluation-set labels. 
In this work, we systematically evaluate how training gender composition shapes fairness in audio deepfake detection. Our main contributions are:

\begin{enumerate}

\item We present the first study (to our knowledge) that systematically varies training gender composition as a controlled independent variable in audio deepfake detection. We train 384 attack-specific models (288 WavLM-Base+ across 9 configurations; 96 LogSpectrogram across 3 configurations) and measure the resulting changes in six fairness metrics. We validate all findings with bootstrap-based significance tests and Benjamini--Hochberg correction~\cite{benjamini1995fdr}.

\item We show that the underrepresented gender in training suffers higher error rates at test time in 31 of 32 attacks, regardless of which gender occupies the minority position. We further show that combined training reduces the mean EER gap from 1.042\,pp to 0.273\,pp compared to within-attack balancing at a similar gender ratio.
% Note that these two configurations differ in both data volume and pooling scope; Section~\ref{subsubsec:same_attack_fairness} isolates the pooling effect directly.

\item We compare LogSpectrogram and WavLM-Base+ and show that they exhibit systematically different bias patterns on identical training data. We show that balanced training nearly eliminates LogSpectrogram disparity but leaves WavLM disparity largely intact, indicating that gender correlations embedded during SSL pre-training persist through fine-tuning regardless of composition~\cite{chen2022wavlm, das2026odyssey}.

\item We demonstrate that voice conversion attacks produce the largest gender gaps under same-attack evaluation by preserving source speaker gender through the conversion process~\cite{c21}. We also find that adversarial perturbation attacks produce the smallest EER gap but the largest demographic parity disparity under same-attack evaluation, making them simultaneously the most and least fair depending on the metric~\cite{c22, hardt2016equality}. These results show that a single metric cannot capture both dimensions.

\item We show that all six calibration strategies, including Oracle calibration, leave the EER gap unchanged at 1.317\,pp across all nine configurations. We attribute this to the fact that threshold adjustment shifts the operating point along the ROC curve but cannot alter its shape; distribution-level EER disparities therefore require retraining, not recalibration. We further show that per-gender EER calibration~\cite{fursule2026smc}, which reduced disparity by 54 to 75\% under balanced training, worsens equalized odds under imbalanced training, a failure mode invisible in single-composition experiments.
\end{enumerate}

Together, these results show that fairness disparities in audio deepfake detection are not merely artifacts of threshold selection, but arise from the interaction between training composition, attack type, and feature representation. This finding reframes fairness mitigation in spoofing countermeasures as a data- and representation-level problem rather than a purely post-hoc calibration problem. 

The remainder of this paper is organized to support this central argument. Section~\ref{sec:related} reviews related work on audio deepfake detection, fairness evaluation, and demographic disparities in spoofing countermeasures. Section~\ref{sec:data} describes the experimental setup, including the dataset, training configurations, feature representations, classifier architecture, evaluation protocol, fairness metrics, threshold calibration strategies, and statistical validation procedure. Section~\ref{sec:results} presents and analyzes the experimental results across training configurations, attack types, feature representations, and threshold calibration strategies. Section~\ref{sec:limitations} discusses the limitations of this study, and Section~\ref{sec:conclusion} concludes the paper.

\begin{table*}[!t]
\centering
\caption{Spoofing attacks in ASVspoof5~\cite{c4} across TTS 
(text-to-speech), VC (voice conversion), and AT 
(adversarial perturbation). F (Female) and M (Male) columns 
show total number of utterances per gender across all splits.}
\label{tab:attacks}
\resizebox{\textwidth}{!}{%
\footnotesize
\begin{tabular}{|c|l|l|r|r|c|l|l|r|r|}
\hline
ID & Type & Algorithm 
& F & M
& ID & Type & Algorithm 
& F & M \\
\hline
A01 & TTS & GlowTTS \cite{kim2020glowtts}         & 10095 & 10350 & A17 & TTS & ZMM-TTS \cite{gong2024zmmtts}        & 8660 & 7497 \\
A02 & TTS & Variant of A01                         & 10095 & 10350 & A18 & AT  & A17+Malafide                          & 8508 & 7411 \\
A03 & TTS & Variant of A01                         & 10095 & 10350 & A19 & TTS & MaryTTS \cite{steiner2018marytts}     & 7430 & 6773 \\
A04 & TTS & GradTTS \cite{popov2021gradtts}        & 10095 & 10350 & A20 & AT  & A12+Malafide                          & 7694 & 6910 \\
A05 & TTS & Variant of A04                         & 10095 & 10350 & A21 & TTS & A09+BigVGAN \cite{lee2022bigvgan}     & 8513 & 7478 \\
A06 & TTS & Variant of A04                         & 10095 & 10350 & A22 & TTS & Variant of A09 \cite{lux2023prosody}  & 8350 & 7491 \\
A07 & TTS & FastPitch \cite{lancucki2021fastpitch}  & 10095 & 10350 & A23 & AT  & A09+Malafide                          & 8389 & 7438 \\
A08 & TTS & VITS \cite{kim2021vits}                & 10095 & 10350 & A24 & VC  & In-house ASR                          & 8533 & 7463 \\
A09 & TTS & ToucanTTS \cite{lux2022lowresource}    &  6695 &  7007 & A25 & VC  & DiffVC \cite{popov2021diffvc}         & 8484 & 7519 \\
A10 & TTS & A09+HiFiGAN \cite{kong2020hifigan}     &  6695 &  7007 & A26 & VC  & A16+genuine noise                     & 8472 & 7522 \\
A11 & TTS & Tacotron2 \cite{shen2018tacotron2}     &  6695 &  7007 & A27 & AT  & A26+Malacopula                        & 8549 & 7355 \\
A12 & TTS & Unit-select                             &  6695 &  7007 & A28 & TTS & YourTTS \cite{casanova2022yourtts}    & 8454 & 7507 \\
A13 & VC  & StarGANv2-VC \cite{li2021starganv2vc}  &  6695 &  7007 & A29 & TTS & XTTS \cite{casanova2024xtts}          & 8395 & 7549 \\
A14 & TTS & YourTTS \cite{casanova2022yourtts}     &  6695 &  7007 & A30 & AT  & A18+Mal.+Malacop.                     & 8405 & 7466 \\
A15 & VC  & VAE-GAN \cite{albadawy2020voice}       &  6695 &  7007 & A31 & AT  & A22+Malacopula                        & 8285 & 7418 \\
A16 & VC  & In-house ASR                           &  6695 &  7007 & A32 & AT  & A25+Malacopula                        & 8194 & 7314 \\
\hline
\end{tabular}%
}
\end{table*}

\section{Related Work}
\label{sec:related}

\subsection{Audio Deepfake Detection}
\label{subsec:rw_detection}

Detection methods for synthetic speech have evolved from classical Gaussian mixture model classifiers to end-to-end architectures such as RawNet2~\cite{c6}, graph attention networks such as AASIST~\cite{c7}, and self-supervised representations such as WavLM~\cite{chen2022wavlm}. The ASVspoof challenge series~\cite{wu2017asvspoof, yamagishi2021asvspoof, liu2023asvspoof, c4} has driven this progress by providing standardized benchmarks across neural TTS, voice conversion, and adversarial perturbation scenarios. Pre-trained large-scale TTS systems such as NaturalSpeech~\cite{tan2024naturalspeech} now generate highly natural speech with speaker-specific prosodic and acoustic characteristics, increasing the difficulty of detecting synthetic speech when demographic cues are preserved in the generated signal. The PhonemeDF dataset~\cite{c30} and phonetic analyses using HuBERT embeddings~\cite{c29, nallaguntla2026phoneme} further characterize the acoustic properties of real and synthetic speech that detectors rely on, including gender-correlated spectral features. Studies on SSL frontend generalization show that SSL representations trained on large-scale corpora do not generalize equally to all speaker subgroups~\cite{das2026odyssey}, suggesting that demographic disparities in detection performance may originate partly in the pre-training stage rather than in fine-tuning or architecture choice. Despite strong average performance, detection systems show systematic demographic disparities that aggregate metrics do not expose, motivating the fairness analysis in this paper.

\subsection{Demographic Bias in Audio Deepfake Detection}
\label{subsec:rw_gender}

Fairness in speaker verification and audio deepfake detection has received growing attention, with studies consistently documenting performance gaps across demographic groups~\cite{hutiri2022bias}. An early analysis of speaker recognition~\cite{c14} showed that models trained on gender-imbalanced corpora yield unequal verification performance for male and female speakers. Research on adversarial and multi-task strategies for speaker verification~\cite{peri2022adversarial} found that adversarial debiasing reduces gender gaps but also reduces overall system utility, leading practitioners to prefer post-hoc solutions that do not require retraining. Empirical work on self-supervised learning~\cite{c19} shows that pre-training data distribution can strongly influence the representations a model encodes, and that such information may persist through downstream fine-tuning.

In audio deepfake detection specifically, gender-dependent performance has been observed from several directions. One study~\cite{c15} found that detectors trained on female speech outperform those trained on male speech, attributing this to high-pitched spectral artifacts in synthetic female voices. A structured fairness audit across six detection systems~\cite{yadav2024fairssd} showed that male speakers consistently face higher false positive rates regardless of architecture. Independent work~\cite{c21} confirmed that score distribution shifts between genders contribute to bias independently of spectral artifacts, and proposed artifact-focused self-synthesis as a training-time mitigation strategy; this approach is complementary to the composition-controlled analysis in this study, which evaluates post-hoc calibration rather than training-time intervention. A gender-balanced deepfake dataset~\cite{c25} and the SCDF dataset~\cite{stanvek2025scdf}, which provides speaker characteristics including gender, age, and dialect annotations to enable demographic fairness evaluation, reflect the community's recognition of gender imbalance as a structural problem requiring dedicated evaluation resources. Fairness analysis in video deepfake detection~\cite{trinh2021examination, xu2024analyzing,c26, c27} confirm that demographic disparities persist across modalities and that balanced training does not fully eliminate them. A study on speech enhancement under noisy conditions~\cite{c31} further motivates robust fairness evaluation under diverse deployment scenarios beyond clean laboratory conditions.

Most directly related to this work, \cite{c18} conducted the first comprehensive gender fairness evaluation on ASVspoof5 using five group fairness metrics under a fixed training composition, and \cite{fursule2026smc} showed that per-gender threshold calibration reduces false alarm disparity by 54 to 75\% under balanced training. Neither study varied training gender composition, leaving it unknown whether bias direction is predictable from composition or whether the method of achieving gender balance affects fairness outcomes.

\subsection{Fairness Theory and Mitigation Strategies}
\label{subsec:rw_fairness}

The machine learning fairness literature provides the theoretical grounding for this study. A comprehensive survey of bias sources across the ML pipeline~\cite{c22} categorized fairness metrics into group, individual, and causal families. A study on how bias enters AI systems at the data collection, representation, and training stages~\cite{c13} shows that data-stage decisions produce measurable fairness consequences at test time. An intervention-based analysis of spoofing countermeasures~\cite{rubio2026odyssey} further shows that data-stage decisions create confounded statistical pathways that persist at inference time regardless of model scale. Empirical work on standard empirical risk minimization~\cite{c19} showed that minority groups experience disproportionate performance degradation as training progresses, even when overall accuracy improves.

The foundational equalized odds criterion~\cite{hardt2016equality} requires model outputs to be conditionally independent of group membership given the true label. The impossibility result in~\cite{chouldechova2017fair} shows that equalized odds, predictive parity, and calibration cannot simultaneously hold when base rates differ across groups. These theoretical constraints predict that no single post-hoc calibration strategy can satisfy all fairness criteria simultaneously. Post-processing methods attract practitioners because they require no retraining~\cite{c22}. The equalized odds framework~\cite{hardt2016equality} provides a principled post-hoc approach, and per-gender threshold calibration has been applied in speaker verification to reduce demographic disparities~\cite{peri2022adversarial}, though its behavior under varied training compositions in deepfake detection has not been characterized.

To the best of our knowledge, no prior study has varied training gender composition as a controlled experimental variable, evaluated multi-strategy calibration across diverse compositions, or formally characterized where post-hoc correction reaches a hard boundary. This paper addresses all three gaps across 384 models, nine configurations, 32 attacks, and six calibration strategies.

 %============================================================
\section{Experimental Setup}
\label{sec:data}
\begin{figure*}[!t]
\centering
\includegraphics[width=\linewidth]{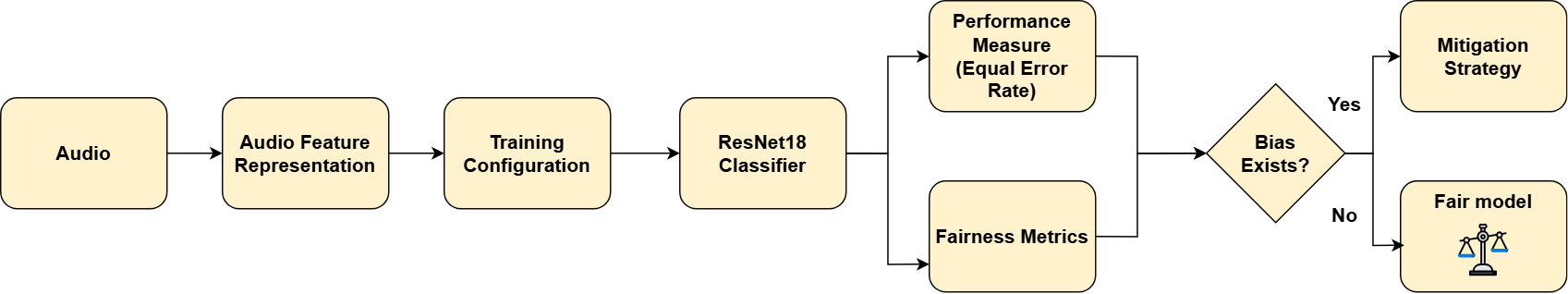}
\caption{Overview of the proposed system for gender fairness evaluation and post-hoc mitigation in audio deepfake detection.}
\label{fig:methodology_pipeline}
\end{figure*}

In this section, we describe the dataset, 
gender-composition configurations, system architecture, 
fairness metrics, and statistical validation procedure 
used in this study. Fig.~\ref{fig:methodology_pipeline} 
summarizes the overall system. Raw audio passes through 
two front-end feature representations into a shared 
ResNet18 classifier backend, trained under nine 
gender-composition configurations. We evaluate each model 
using EER and six group fairness metrics, and we apply 
six post-hoc threshold calibration strategies to assess 
how far fairness can be improved without retraining.

\begin{table}[!t]
\centering
\caption{Summary of training, development, and evaluation splits pooled across
all attacks by gender and class.}
\label{tab:split_summary}
\renewcommand{\arraystretch}{1.12}
\setlength{\tabcolsep}{8pt}
\small
\resizebox{\columnwidth}{!}{%
\begin{tabular}{|l|c|c|c|c|}
\hline
\multirow{2}{*}{Split}
& \multicolumn{2}{c|}{Female}
& \multicolumn{2}{c|}{Male} \\
\cline{2-5}
& Spoof & Bonafide
& Spoof & Bonafide\\
\hline
Train & 187330 & 32034 & 179854 & 40022 \\
Dev   &  26749 &  4576 &  25683 &  5717 \\
Eval  &  53556 &  9154 &  51430 & 11436 \\
\hline
\end{tabular}%
}
\end{table}

\subsection{ASVspoof5 Dataset}
\label{subsec:dataset}
All experiments use ASVspoof5~\cite{c4}, the most recent ASVspoof challenge benchmark. It covers 32 distinct spoofing systems (A01--A32) spanning neural TTS, voice conversion (VC), and adversarial perturbation (AT) attacks, making it well-suited for studying attack-dependent fairness. The dataset is divided into three official partitions: training (A01--A08), development (A09--A16), and evaluation (A17--A32) each containing bonafide utterances denoted as Bonafide Train (BT), Bonafide Dev (BD), and Bonafide Eval (BE) respectively, alongside spoofed speech spanning neural TTS, voice conversion, and adversarial perturbation attacks, whose attack identifiers and algorithms are listed in Table~\ref{tab:attacks}. 
 
Both female and male speakers appear across all 32 attacks. The training partition contains approximately 10,000 utterances per gender per attack; the development partition contains 6,700--6,900; the evaluation partition contains approximately 18,000 female and 16,000 male utterances per attack. This reflects a slight female-majority imbalance (53\%\,F\,/\,47\%\,M) inherent to ASVspoof5 and not to our experimental design. Gender labels and class annotations come directly from the official ASVspoof5 protocol files. All audio signals are resampled to 16\,kHz; feature-specific duration normalization is described in Section~\ref{subsec:features}, following an approach similar to that in~\cite{pascu2025easy}.

\begin{figure*}[t]
\centering
\includegraphics[width=\textwidth]{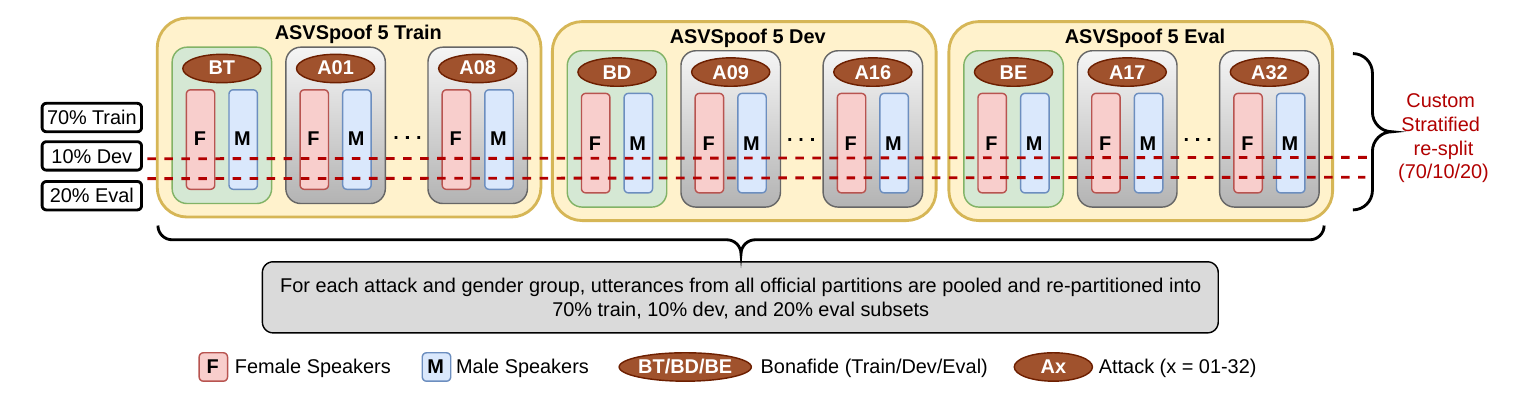}
\caption{ASVspoof5 dataset protocol used in this study.  (A01-A08 training, 
A09 -A16 development, A17-A32 evaluation).}
\label{fig:dataset}
\end{figure*}

\subsection{Gender-Composition Training Configurations}
\label{subsec:configs}

Rather than following the official ASVspoof5 partition, we construct custom per-gender per-attack splits using all utterances from all 32 attacks regardless of their official partition assignment. For each (attack, gender) pair, all available utterances are pooled and divided into a stratified 70\,/\,10\,/\,20\% split for training, development, and evaluation respectively, ensuring every attack appears in every split and gender proportions are controlled explicitly, as illustrated in Fig. \ref{fig:dataset}. Table \ref{tab:split_summary} reports the resulting utterance counts pooled across all 32 attacks, by split, gender, and class. Splits are stratified by attack and gender but not guaranteed to be speaker-disjoint; potential speaker overlap across splits is acknowledged as a limitation in Section~\ref{sec:limitations}. We confirm gender balance in our custom training split via chi-squared test ($\chi^2 = 0.597$, $p = 0.440$), ruling out training-set gender imbalance as a confound for the composition experiments. The development set serves two purposes: monitoring validation loss for early stopping during training, and calibrating the decision threshold at the global EER operating point before evaluation. From this pool we construct nine gender-composition configurations (Table~\ref{tab:configs}). F+$X$\%M denotes a female-anchored configuration using 100\% of available female utterances plus $X$\% of available male utterances added to the training set. For example, F+25\%M uses all female utterances and adds 25\% of male utterances, making female speakers the dominant group. M+$X$\%F denotes the symmetric male-anchored case: 100\% of available male utterances plus $X$\% of available female utterances. An internal validity check is embedded in this design.  F+75\%M uses all female utterances plus 75\% of male utterances; M+25\%F uses all male utterances plus only 25\% of female utterances. Both produce a training set where one gender contributes 100\% of its pool and the other contributes 25\%, but the dominant gender differs. If both configurations produce the same bias direction despite using entirely different utterance pools, the dominant gender in training drives bias, not the specific utterances selected. Section~\ref{subsubsec:same_attack_fairness} confirms this. 

The nine configurations form two experimental tiers. Baseline (F, M, Combined) uses the full pool of one gender or both equally, evaluated under both LogSpectrogram and WavLM-Base+ to isolate how feature type modulates bias. GenderMix (the six partial-mix configurations) uses WavLM-Base+ only, because WavLM exhibits substantially larger and more widespread gender disparities than LogSpectrogram in baseline experiments (Section~\ref{subsec:baseline}), making it the more consequential representation for studying composition effects at scale. Together the nine configurations span the full spectrum from gender-exclusive to balanced, enabling a systematic controlled analysis of training composition as a fairness variable.

\begin{table}[!t]
\centering
\caption{Gender-composition training configurations for F (Female) and M (Male). F+$X$\%M denotes
female-anchored training with $X$\% male utterances added; M+$X$\%F denotes
the symmetric male-anchored case.}
\label{tab:configs}
\renewcommand{\arraystretch}{1.12}
\setlength{\tabcolsep}{6pt}
% \scriptsize
% \resizebox{\columnwidth}{!}{%
\small
\resizebox{\columnwidth}{!}{%
\begin{tabular}{|l|c|c|}
\hline
Config & F pool used (\%) & M pool used (\%) \\
\hline
\multicolumn{3}{|l|}{\textit{Baseline}} \\
\hline
F (Female only)       & 100 & 0   \\
M  (Male only)      & 0   & 100 \\
Combined & 100 & 100 \\
\hline
\multicolumn{3}{|l|}{\textit{Female-anchored GenderMix}} \\
\hline
F+25\%M  & 100 & 25  \\
F+50\%M  & 100 & 50  \\
F+75\%M  & 100 & 75  \\
\hline
\multicolumn{3}{|l|}{\textit{Male-anchored GenderMix}} \\
\hline
M+75\%F  & 75  & 100 \\
M+50\%F  & 50  & 100 \\
M+25\%F  & 25  & 100 \\
\hline
\end{tabular}%
}
\end{table}

\subsection{Feature Representations}
\label{subsec:features}
%-------------------------------------------------------------

We resample all audio signals to 16\,kHz prior to feature 
extraction. Duration normalization differs by feature and 
is described in each subsection below. Baseline 
configurations use both LogSpectrogram and WavLM-Base+; 
GenderMix configurations use WavLM-Base+ only. We select 
these two representations to contrast a handcrafted 
time-frequency feature whose bias traces to vocoder 
spectral artifacts with a pre-trained SSL model whose 
gender encoding originates upstream of fine-tuning, 
independently of training 
composition~\cite{hanilci2026cyclostationarity}.

\subsubsection{LogSpectrogram}
\label{subsubsec:logspec}

The LogSpectrogram (LogSpec) applies a logarithmic 
transformation to the magnitude spectrogram, capturing 
frequency-band energy variations informative for 
detecting artifacts introduced by speech synthesis and 
voice conversion~\cite{c18}. We zero-pad signals shorter 
than 4.0\,s and truncate signals longer than 4.0\,s. We 
compute features with FFT size 800, hop length 320 
samples, and a Hann window, yielding a feature map of 
shape $(1 \times 401 \times 201)$ corresponding to 401 
frequency bins and 201 time frames, which the ResNet18 
classifier treats as a single-channel image. We evaluate 
LogSpec under the three baseline configurations only (F, 
M, Combined), providing a diagnostic reference for how 
feature type modulates bias before the WavLM-based 
GenderMix analysis.

\subsubsection{WavLM-Base+}
\label{subsubsec:wavlm}

WavLM-Base+~\cite{chen2022wavlm} is a self-supervised 
speech model pre-trained on 94,000 hours of unlabeled 
speech using a masked speech prediction objective with 
denoising. It consists of a CNN feature encoder followed 
by 12 Transformer layers producing contextualized 
frame-level embeddings of dimension 768 at a stride of 
320 samples (20\,ms at 16\,kHz). WavLM embeddings encode 
detailed speaker-level information, including pitch 
contours, vocal tract resonances, and speaking style. We 
hypothesize that this encoding contributes to 
gender-differential detection performance across attack 
types because these properties differ systematically 
between male and female speakers. Rather than 
zero-padding, we repeat signals cyclically to exactly 
64,600 samples ($\approx$4.04\,s) to avoid introducing 
silence regions that would alter the SSL feature 
statistics. We extract representations from the final 
Transformer layer, which captures speaker-sensitive 
identity and demographic cues~\cite{chen2022wavlm}, 
yielding a fixed-size feature map of shape 
$(201 \times 768)$ where 
$T = \lfloor 64600 / 320 \rfloor = 201$ frames. We 
pre-extract all features offline prior to model training, 
and all 288 WavLM models load these identical 
pre-computed features, ensuring that observed differences 
between configurations reflect only the 
gender-composition variable.

%-------------------------------------------------------------
\subsection{ResNet18 Classifier Backend}
\label{subsec:classifier}
%-------------------------------------------------------------

Both feature representations pass through a ResNet18 
architecture~\cite{he2016resnet} with a binary output 
head. We treat the 2D feature map as a single-channel 
image, pass it through the ResNet18 convolutional trunk, 
and apply adaptive average pooling followed by a linear 
layer producing a two-class logit. This backend matches 
the design in~\cite{c18}, which used the same ResNet18 
architecture for fairness evaluation on ASVspoof5, 
ensuring that fairness differences in this study are 
attributable to feature representation and training 
composition rather than classifier architecture.

We train each model independently using 
AdamW~\cite{loshchilov2019adamw} with learning rate 
$3 \times 10^{-5}$ and weight decay $10^{-4}$. To 
account for class imbalance between bonafide and spoofed 
utterances within each attack-specific training set, we 
apply class-weighted binary cross-entropy:

\begin{equation}
w_c = \frac{N}{C \cdot N_c}, \quad
c \in \{\text{bonafide, spoof}\}
\label{eq:class_weights}
\end{equation}

\noindent where $N$ is total training samples, $C{=}2$, 
and $N_c$ is the count of class $c$. We train each model 
for a maximum of 100 epochs with early stopping on 
minimum development loss (patience\,=\,15 epochs), and 
we apply a ReduceLROnPlateau 
scheduler~\cite{pytorch2019} that halves the learning 
rate when development loss does not improve for 5 
consecutive epochs. We use the checkpoint achieving 
minimum development loss for all evaluations.

To ensure reproducibility, we determine the 
gender-composition subset for each configuration prior to 
training using fixed random seed $s{=}42$, ensuring 
identical utterance pools across all experimental runs. 
Mini-batches use per-epoch shuffling during training; 
development and evaluation loaders use sequential 
loading. This strict control over randomness ensures that 
observed differences between configurations reflect only 
the gender-composition variable and not variation in file 
selection or batch ordering.

%-------------------------------------------------------------
\subsection{Evaluation Protocol}
\label{subsec:protocol}
%-------------------------------------------------------------

We train one attack-specific model per (attack, 
configuration) pair: 288 WavLM-Base+ models across all 
nine configurations (9\,$\times$\,32), plus 96 
LogSpectrogram models for the three baseline 
configurations (3\,$\times$\,32), yielding 384 models in 
total. Each model trains solely on the attack it targets, 
following the per-attack paradigm that isolates 
attack-specific bias from cross-attack generalization 
effects~\cite{mohamed2022selfsupervised, 
sen2025noiseaware, zhang2022c3dino}. At evaluation time, 
we run each model separately on the female and male 
subsets of the held-out evaluation pool, yielding 
per-gender EER estimates~\cite{sen2025noiseaware, 
gao2026context}. We calibrate the decision threshold on 
the development set at the global EER operating point and 
keep it fixed during evaluation, ensuring gender 
differences reflect score distribution disparities rather 
than threshold placement~\cite{c18, gao2026context}.

%-------------------------------------------------------------
\subsection{Fairness Evaluation Framework}
\label{sec:framework}
%-------------------------------------------------------------

Here, we formalize the metrics used to quantify detection 
performance and gender fairness, the threshold 
calibration strategies evaluated as post-hoc mitigation, 
and the statistical validation procedure applied 
throughout. Let $G \in \{f, m\}$ denote speaker gender, 
$Y \in \{0,1\}$ the true label (0\,=\,bonafide, 
1\,=\,spoof), and $\hat{Y} = \mathbf{1}[s \geq \theta] 
\in \{0,1\}$ the model prediction, where $s$ is the 
detector's spoof score and $\theta$ is the decision 
threshold. $\text{TP}_g$, $\text{TN}_g$, $\text{FP}_g$, 
$\text{FN}_g$ denote confusion matrix counts restricted 
to gender group $g \in \{f, m\}$.

\subsubsection{Detection Performance Metric}
\label{subsec:perf_metrics}

The Equal Error Rate (EER) marks the operating point 
where the false alarm rate and miss rate are equal:

\begin{equation}
    \text{EER}:\ \text{FAR}(\theta) = 
    \text{FRR}(\theta)
    \label{eq:eer}
\end{equation}

We compute per-gender EER values $\text{EER}_f$ and 
$\text{EER}_m$ by finding, for each gender independently, 
the threshold at which that gender's false alarm rate 
equals its miss rate. We then measure gender disparity 
in detection performance as the absolute EER gap:

\begin{equation}
    \Delta\text{EER} = \left|\,\text{EER}_f -
    \text{EER}_m\right|
    \label{eq:eer_gap}
\end{equation}

We use the absolute gap rather than a relative ratio 
because a ratio is undefined when either gender achieves 
near-zero EER, which occurs for several attacks in this 
study.

\subsubsection{Group Fairness Metrics}
\label{subsec:fairness_metrics}

We evaluate six group fairness metrics that collectively 
capture distinct reliability failure modes across 
different deployment contexts. Each metric equals zero 
when the model treats both genders identically on the 
corresponding criterion. We report all six because they 
capture distinct fairness dimensions that no single 
metric can summarize, as the divergent attack rankings 
in Section~\ref{sec:results} illustrate.

SPD measures whether both genders receive equal 
spoof-flagging rates regardless of ground 
truth~\cite{dwork2012fairness}:

\begin{equation}
    \text{SPD} = \left|P(\hat{Y}{=}1 \mid G{=}f) -
                       P(\hat{Y}{=}1 \mid G{=}m)
                       \right|
    \label{eq:spd}
\end{equation}

A high SPD means one gender is flagged as spoof more 
frequently than the other, independent of whether the 
flagging is correct. SPD is the relevant criterion for 
broadcast monitoring and content moderation, where the 
demographic reach of flagging decisions must be 
equitable.

EOpD measures whether spoofed utterances are detected at 
equal rates for both genders~\cite{hardt2016equality}:

\begin{equation}
    \text{EOpD} = \left|
        \frac{\text{TP}_f}{\text{TP}_f + \text{FN}_f}
        -
        \frac{\text{TP}_m}{\text{TP}_m + \text{FN}_m}
    \right|
    \label{eq:eopd}
\end{equation}

A high EOpD means the model misses more spoof utterances 
from one gender than the other and is the primary 
criterion for contexts where missed detections are the 
critical failure mode, such as fraud detection and 
identity verification.

EOD measures combined disparity in both true positive 
and false positive rates across 
genders~\cite{pessach2022review}:

\begin{equation}
\begin{aligned}
    \text{EOD} =
    &\left|
        \frac{\text{TP}_f}{\text{TP}_f + \text{FN}_f}
        -
        \frac{\text{TP}_m}{\text{TP}_m + \text{FN}_m}
    \right| \\
    &+
    \left|
        \frac{\text{FP}_f}{\text{FP}_f + \text{TN}_f}
        -
        \frac{\text{FP}_m}{\text{FP}_m + \text{TN}_m}
    \right|
\end{aligned}
\label{eq:eod}
\end{equation}

EOD equals zero only when the model simultaneously 
equalizes detection rates and false alarm rates across 
genders, making it a stricter criterion than EOpD alone 
and relevant when both missed detections and false alarms 
carry significant consequences.

$\text{FPR}_\text{gap}$ measures whether bonafide 
speakers receive equal false rejection rates across 
genders~\cite{verma2018fairness}:

\begin{equation}
    \text{FPR}_\text{gap} =
    \left|
        \frac{\text{FP}_f}{\text{FP}_f + \text{TN}_f}
        -
        \frac{\text{FP}_m}{\text{FP}_m + \text{TN}_m}
    \right|
    \label{eq:fpr_gap}
\end{equation}

A high $\text{FPR}_\text{gap}$ means bonafide speakers 
of one gender are more frequently rejected as spoof than 
the other, directly affecting user experience equity in 
voice authentication systems.

PPD measures whether spoof predictions carry equal 
precision across genders~\cite{hellman2020measuring}:

\begin{equation}
    \text{PPD} =
    \left|
        \frac{\text{TP}_f}{\text{TP}_f + \text{FP}_f}
        -
        \frac{\text{TP}_m}{\text{TP}_m + \text{FP}_m}
    \right|
    \label{eq:ppd}
\end{equation}

A high PPD means spoof predictions are less reliable for 
one gender than the other and is relevant in legal or 
forensic applications where prediction precision directly 
determines downstream decisions.

TED measures asymmetry in the ratio of false alarms to 
missed detections across genders~\cite{wachter2021bias}:

\begin{equation}
    \text{TED} =
    \left|
        \frac{\text{FP}_f}{\text{FN}_f} -
        \frac{\text{FP}_m}{\text{FN}_m}
    \right|
    \label{eq:ted}
\end{equation}

TED\,=\,0 means the model makes proportionally the same 
mix of false alarms and missed detections for both 
genders. A large TED indicates the model primarily 
false-alarms one gender while primarily missing spoof 
detections for the other, a failure mode not captured by 
rate-based metrics. We exclude cases where 
$\text{FN}_g = 0$ for either gender, as TED is undefined 
in those cases. TED captures a distinct fairness 
dimension from the five rate-based metrics above, as the 
correlation analysis in 
Section~\ref{subsec:disc_deployment} confirms.

\begin{table*}[!t]
\centering
\caption{Feature comparison in terms of mean EER over 32 same-attack evaluations. F\_EER and M\_EER denote female and male EER, respectively. Gap = $|$F\_EER $-$ M\_EER$|$.}
\label{tab:baseline}
\renewcommand{\arraystretch}{1.2}
\setlength{\tabcolsep}{3pt}
\small
\begin{tabular*}{\textwidth}{@{\extracolsep{\fill}}|c|c|c|c|c|c|c|c|c|}
\hline
\multirow{2}{*}{Config}
& \multicolumn{4}{c|}{LogSpectrogram}
& \multicolumn{4}{c|}{WavLM-Base+} \\
\cline{2-9}
& F\_EER
& M\_EER
& EER gap
& Biased
& F\_EER
& M\_EER
& EER gap
& Biased \\
\hline
F
& 0.133 & 0.768 & 0.635 & 10/32
& 2.782 & 4.690 & 1.917 & 32/32 \\
M
& 0.665 & 0.315 & 0.621 & 9/32
& 5.056 & 2.780 & 2.276 & 32/32 \\
Combined
& 0.095 & 0.156 & \textbf{0.063} & \textbf{2/32}
& 2.226 & 2.180 & \textbf{0.273} & \textbf{20/32} \\
\hline
\end{tabular*}
\end{table*}

\begin{table*}[!t]
\centering
\caption{Feature comparison in terms of mean EER over 32 Cross-Attack). F\_EER (Female EER) and M\_ EER (Male EER), Gap = $|$F\_EER $-$ M\_EER$|$}
\label{tab:baseline_cross}
\renewcommand{\arraystretch}{1.2}
\setlength{\tabcolsep}{3pt}
\small
\begin{tabular*}{\textwidth}{@{\extracolsep{\fill}}
|c|c|c|c|c|c|c|c|c|}
\hline
\multirow{2}{*}{Config}
& \multicolumn{4}{c|}{LogSpectrogram}
& \multicolumn{4}{c|}{WavLM-Base+} \\
\cline{2-9}
& F\_EER
& M\_EER
& EER gap
& Biased
& F\_EER
& M\_EER
& EER gap
& Biased \\
\hline
F
& 33.207 & 34.151 & 0.944 & 32/32
& 30.423 & 30.803 & 1.591 & 32/32 \\
M
& 34.292 & 33.492 & 0.800 & 32/32
& 31.905 & 30.552 & 1.963 & 32/32 \\
Combined
& 33.158 & 32.953 & \textbf{0.205} & 32/32
& 30.918 & 30.464 & \textbf{1.491} & 32/32 \\
\hline
\end{tabular*}
\end{table*}

\subsubsection{Threshold Calibration Strategies}
\label{subsec:tc_strategies}

We evaluate six threshold calibration strategies as 
post-hoc mitigation for improving group fairness without 
retraining. Each detector produces a spoof score $s$; a 
higher score indicates stronger evidence that the 
utterance is spoofed. We select thresholds by 
interpolating the development-set ROC curve to find the 
exact operating point satisfying each strategy's 
criterion, and we apply them unchanged to the held-out 
evaluation set.

TC0 sets a single global threshold $\theta^*$ at the 
operating point where the combined false alarm rate 
equals the combined miss rate across both genders, 
without gender-specific adjustment, and serves as the 
standard deployment baseline. TC\_EER sets separate 
per-gender thresholds $\theta_f^*$ and $\theta_m^*$, 
each at the per-gender EER operating point on the 
development set, following the gender-specific 
thresholding approach of~\cite{jain2021inclusive} and 
as used in our prior work~\cite{fursule2026smc}. TC\_FPR 
sets per-gender thresholds such that 
$\text{FPR}_f(\theta_f) = \text{FPR}_m(\theta_m)$ on 
the development set, targeting 
$\text{FPR}_\text{gap}\,\approx\,0$~\cite{pleiss2017fairness}. 
TC\_TPR sets per-gender thresholds such that 
$\text{TPR}_f(\theta_f) = \text{TPR}_m(\theta_m)$ on 
the development set, targeting 
$\text{EOpD}\,\approx\,0$~\cite{pleiss2017fairness}. 
TC\_DP sets per-gender thresholds such that the overall 
positive prediction rate is equal across genders on the 
development set, targeting 
$\text{SPD}\,\approx\,0$~\cite{dwork2012fairness}. 
Finally, Oracle selects per-gender thresholds using 
evaluation-set labels directly; this strategy is not 
deployable and serves only as the theoretical upper 
bound on what any threshold-based post-hoc strategy can 
achieve~\cite{hardt2016equality, pleiss2017fairness}.

We note that threshold calibration changes the operating 
point of a detector but does not change the underlying 
score distributions. Selecting a different threshold can 
change deployment-time FPR, TPR, SPD, or EOD, but 
cannot reshape the ROC curve or eliminate 
distribution-level EER disparities. We therefore treat 
EER gap as the primary bias indicator throughout this 
study.

%-------------------------------------------------------------
\subsection{Statistical Significance and Multiple 
Comparison Correction}
\label{subsec:significance}
%-------------------------------------------------------------

We validate all reported fairness gaps via Poisson 
bootstrap resampling~\cite{hanley2006bootstrap} with 
$N{=}1000$ resamples per model. For each resample, we 
draw confusion matrix counts independently from Poisson 
distributions parameterized by the observed counts, 
treating each cell as an independent count rather than 
resampling individual utterances. This approximation is 
valid when utterance-level correlations between confusion 
matrix cells are small relative to cell magnitudes, which 
holds for the evaluation set sizes used in this 
study~\cite{hanley2006bootstrap}. We estimate bootstrap 
p-values as the proportion of resamples in which the 
resampled metric value equals zero, consistent with 
testing $H_0\text{: gap\,=\,0}$ against a two-sided 
alternative. We acknowledge that $N{=}1000$ resamples 
represents a practical lower bound for significance 
testing at $p < 0.001$, and we identify increasing $N$ 
to 5000 or 10000 as a direction for future 
work~\cite{efron1994introduction}.

We apply multiple comparison correction using the 
Benjamini--Hochberg false discovery rate 
procedure~\cite{benjamini1995fdr} at $\alpha{=}0.05$, 
within each metric across all 288 WavLM-based 
model-level tests (9\,configurations $\times$\,32 
attacks). We apply BH correction within each metric 
rather than across all metrics simultaneously, as 
between-metric dependencies would make joint correction 
overly conservative. BH correction remains valid under 
positive regression dependence among test 
statistics~\cite{benjamini2001fdr}, a condition that 
holds when tests share overlapping utterance pools as 
in this study.

 \begin{figure*}[!t]
\centering
\includegraphics[width=\textwidth]{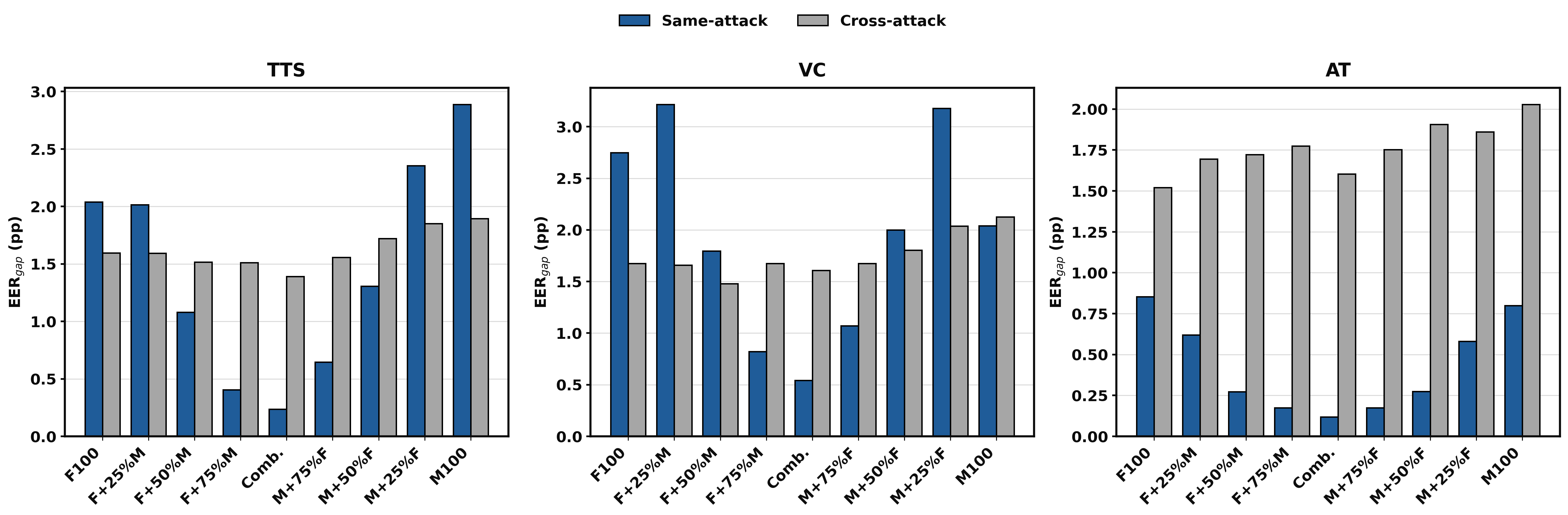}
\caption{EER$_\text{gap}$ under same-attack  and cross-attack evaluation across all nine gender-composition configurations for TTS (text-to-speech), 
VC (voice conversion), and AT (adversarial perturbation) 
attacks.}
\label{fig:same_vs_cross}
\end{figure*}

\section{Results \& Discussion}
\label{sec:results}
 In this section, we present and discuss our experimental results in three parts: feature comparison for gender bias, fairness under varying training composition, and post-hoc mitigation.

%=============================================================
\subsection{Baseline Feature Comparison}
\label{subsec:baseline}
%=============================================================
 
In this first experiment, we investigate whether feature representation itself shapes gender bias, independent of training composition. Table IV addresses this using the three baseline configurations. Under Female-Only training, LogSpectrogram produces an EER gap of 0.635 pp, while WavLM-Base+ produces 1.917 pp (which is three times larger). We observe the same pattern under Male-Only training, where WavLM's gap of 2.276 pp is 3.7 times that of LogSpectrogram's 0.621 pp. Here, we hypothesize that balancing the training data would close the gap for both features in similar fashion. We test this directly using Combined training, which trains both features on exactly the same gender-balanced data and removes composition as a confound. Here, LogSpectrogram bias nearly disappears: it falls to 0.063 pp, with only 2 of 32 attacks remaining biased. WavLM-Base+ behaves very differently. Its gap narrows only modestly, to 0.273 pp, and 20 of 32 attacks remain biased. Rather than converging, the two features diverge further under balanced training: the WavLM-to-LogSpectrogram ratio grows from 3.0x under Female-Only training to 4.3x under Combined training, consistent with \cite{c18}. If balanced data alone explained the disparity, we would expect both features to move toward fairness together. We instead find that only one of them does. These results suggest that the two representations capture different forms of gender information. LogSpectrogram 
captures vocoder spectral artifacts that balanced training 
can neutralize, since those artifacts are tied to the 
attack-specific synthesis process and disappear once 
training data covers both genders equally. WavLM-Base+, 
in contrast, encodes speaker gender as a distributed 
pattern across 768 embedding dimensions during 
self-supervised pre-training, a pattern that fine-tuning 
on balanced data cannot remove~\cite{c19, hutiri2022bias}.

To investigate whether WavLM's resistance to balanced training is specific to the attacks seen during training or reflects a more fundamental property of how the model encodes gender, we evaluate each attack-specific model on cross attacks, as shown in Table \ref{tab:baseline_cross}. We observe that every one of the 32 models, regardless of feature type or configuration, shows a cross-attack EER gap exceeding 0.5 pp. Under Combined training, we find that the EER gap LogSpectrogram had nearly closed under same-attack evaluation reopens only modestly, to 0.205 pp, while WavLM's gap remains essentially unchanged at 1.491 pp. We further find that balanced training reduces LogSpectrogram's cross-attack gap substantially, by 0.739 pp (from 0.944 to 0.205 pp), but reduces WavLM's by only 0.100 pp (from 1.591 to 1.491 pp). Our results indicate that bias direction follows training composition even under these cross conditions. Specifically, Female-Only training consistently produces higher male EER for both features, and Male-Only training reverses this relationship as expected.

Overall, these results indicate that gender 
information is embedded in WavLM-Base+ before fine-tuning 
begins and transfers readily to unseen attacks, making 
training composition an insufficient remedy for its bias. 
LogSpectrogram, by contrast, serves as an informative 
counter-example: its bias responds substantially to 
balanced training in both same-attack and cross-attack 
conditions, while WavLM's does not. Because WavLM-Base+ 
represents the more persistent and therefore more 
practically important case, we use it exclusively for the 
remaining GenderMix and calibration experiments.

\begin{table*}[!t]
\caption{Mean fairness metrics across all nine gender-composition configurations for WavLM-Base+, averaged over 32 same-attack evaluations. All values 
are bootstrap-significant ($p \ll 0.001$, $N=1000$)}
\label{tab:fairness_all9}
\centering
\renewcommand{\arraystretch}{1.2}
\setlength{\tabcolsep}{4pt}
\small
\begin{tabular*}{\textwidth}{@{\extracolsep{\fill}}
    |l|c|c|c|c|c|c|c|c|c|c|}
\hline
Config
& F\_EER
& M\_EER
& EER gap
& FPR\_gap
& SPD
& EOpD
& EOD
& PPD
& TED
& Disadv. \\
& (\%) & (\%) & (pp) & & & & & & & \\
\hline
\multicolumn{11}{|l|}{\textit{Baseline Configurations}} \\
\hline
F
  & 2.782 & 4.690 & 1.917
  & 0.0242 & 0.0204 & 0.0238 & 0.0359 & 0.0039 & 0.115
  & M $\uparrow$ \\

M
  & 5.056 & 2.780 & 2.276
  & 0.0495 & \textbf{0.0105} & 0.0088 & 0.0511 & 0.0089 & 0.363
  & F $\downarrow$ \\
  Combined
  & 2.226 & 2.180 & \textbf{0.273}
  & \textbf{0.0052} & 0.0108 & \textbf{0.0042} & \textbf{0.0061} & \textbf{0.0008} & \textbf{0.098}
  & F $\downarrow$ \\
\hline
\multicolumn{11}{|l|}{\textit{GenderMix — Female-Anchored}} \\
\hline
F+25\%M
  & 1.213 & 3.150 & 1.940
  & 0.0236 & 0.0252 & 0.0167 & 0.0263 & 0.0029 & 0.184
  & M $\uparrow$ \\
F+50\%M
  & 1.133 & 2.174 & 1.042
  & 0.0120 & 0.0253 & 0.0087 & 0.0133 & 0.0013 & 0.100
  & M $\uparrow$ \\
F+75\%M
  & 1.143 & 1.568 & 0.438
  & 0.0050 & 0.0277 & 0.0046 & 0.0068 & 0.0005 & 0.114
  & M $\uparrow$ \\
\hline
\multicolumn{11}{|l|}{\textit{GenderMix — Male-Anchored}} \\
\hline
M+75\%F
  & 1.669 & 1.049 & 0.627
  & 0.0092 & 0.0303 & 0.0027 & 0.0093 & 0.0020 & 0.154
  & F $\downarrow$ \\
M+50\%F
  & 2.315 & 1.100 & 1.215
  & 0.0169 & 0.0321 & 0.0052 & 0.0170 & 0.0034 & 0.154
  & F $\downarrow$ \\
M+25\%F
  & 3.291 & 1.165 & 2.126
  & 0.0323 & 0.0339 & 0.0097 & 0.0324 & 0.0064 & 0.207
  & F $\downarrow$ \\
\hline
\end{tabular*}
\end{table*}

\begin{table*}[!t]
\caption{Mean fairness metrics across all nine gender-composition configurations for WavLM-Base+, averaged over 32 cross-attack evaluations. All values 
are bootstrap-significant ($p \ll 0.001$, $N=1000$)}
\label{tab:fairness_cross_all9}
\centering
\renewcommand{\arraystretch}{1.2}
\setlength{\tabcolsep}{4pt}
\small
\begin{tabular*}{\textwidth}{@{\extracolsep{\fill}}
    |l|c|c|c|c|c|c|c|c|c|c|}
\hline
Config
& F\_EER
& M\_EER
& EER gap
& FPR\_gap
& SPD
& EOpD
& EOD
& PPD
& TED
& Disadv. \\
& (\%) & (\%) & (pp) & & & & & & & \\
\hline
\multicolumn{11}{|l|}{\textit{Baseline Configurations}} \\
\hline
F
  & 30.423 & 30.803 & 1.591
  & 0.0367 & 0.0232 & 0.0238 & 0.0424 & 0.0093 & 54.785
  & M $\uparrow$ \\
M
  & 31.905 & 30.552 & 1.963
  & 0.0346 & 0.0102 & 0.0088 & 0.0371 & 0.0102 & 6.358
  & F $\downarrow$ \\
  Combined
  & 30.918 & 30.464 & \textbf{1.491}
  & \textbf{0.0267} & \textbf{0.0063} & \textbf{0.0042} & \textbf{0.0272} & \textbf{0.0092} & \textbf{19.819}
  & F $\downarrow$ \\
\hline
\multicolumn{11}{|l|}{\textit{GenderMix - Female-Anchored}} \\
\hline
F+25\%M
  & 30.511 & 30.567 & 1.626
  & 0.0392 & 0.0161 & 0.0167 & 0.0426 & 0.0256 & 31.319
  & M $\uparrow$ \\
F+50\%M
  & 30.762 & 30.559 & 1.549
  & 0.0285 & 0.0091 & 0.0087 & 0.0300 & 0.0256 & 28.267
  & F $\downarrow$ \\
F+75\%M
  & 30.858 & 30.556 & 1.597
  & 0.0270 & 0.0073 & 0.0046 & 0.0276 & 0.0258 & 28.667
  & F $\downarrow$ \\
\hline
\multicolumn{11}{|l|}{\textit{GenderMix - Male-Anchored}} \\
\hline
M+75\%F
  & 30.973 & 30.408 & 1.622
  & 0.0272 & 0.0085 & 0.0027 & 0.0276 & 0.0261 & 18.429
  & F $\downarrow$ \\
M+50\%F
  & 31.168 & 30.344 & 1.774
  & 0.0273 & 0.0110 & 0.0052 & 0.0283 & 0.0262 & 17.600
  & F $\downarrow$ \\
M+25\%F
  & 31.390 & 30.293 & 1.885
  & 0.0284 & 0.0150 & 0.0097 & 0.0308 & 0.0265 & 29.526
  & F $\downarrow$ \\
\hline
\end{tabular*}
\end{table*}

%=============================================================
\subsection{Fairness Under Varying Training Composition}
\label{subsec:composition}
%=============================================================
In this section, we investigate the gender bias of WavLM-Base+. We examine how varying the gender composition across nine configurations affects fairness outcomes under both same-attack and cross-attack evaluations. 

\subsubsection{Same-Attack Fairness}
\label{subsubsec:same_attack_fairness}
In this experiment, we investigate how the nine 
gender-composition configurations affect fairness under 
same-attack evaluation. 
Table~\ref{tab:fairness_all9} reports mean fairness metrics across all 32 same-attack evaluations for each of the nine configurations. We find a pattern that holds firmly across all nine: the gender that dominates training ends up advantaged at test time. Female-anchored configurations (F, F+25\%M, F+50\%M, F+75\%M) disadvantage male speakers, while male-anchored configurations reverse this relationship exactly. We observe this behavior in 31 of 32 attacks, across all six fairness metrics simultaneously, with the sole exception of A19 (MaryTTS) under Combined training, where the resulting EER gap does not exceed our statistical significance threshold. Prior work has reported both directions of disparity, sometimes within the same body of literature \cite{c15, yadav2024fairssd}; our results resolve this apparent contradiction by showing that the proportion of speech from each gender during training, rather than any innate acoustic quality of male or female speech, determines which gender a trained model will disadvantage. 

Next, we examine whether this disadvantage can be reduced gradually, or whether it persists until training reaches full balance. We find that increasing the proportion of the minority gender improves the EER gap monotonically, but with diminishing returns. Under female-anchored configurations, supplementing the corpus with minority male speech at 25\%, 50\%, and 75\% reduces the EER gap from 1.940 pp to 1.042 pp to 0.438 pp. We observe the mirror image under male-anchored configurations, where the EER gap falls from 2.126 pp to 1.215 pp to 0.627 pp, in steps of 0.911, 0.588, and 0.354 pp. None of the within-attack GenderMix configurations cross the 0.5 pp threshold we treat as practically meaningful bias; only Combined training, which pools 100\% of both genders across all 32 attacks, reaches 0.273 pp. We attribute the additional improvement from F+75\%M to Combined (0.165 pp) and from M+75\%F to Combined (0.354 pp) to a fairness benefit specific to population-level pooling, one that within-attack balancing alone cannot achieve \cite{c14}.
We further investigate whether all fairness metrics improve simultaneously as the composition becomes more balanced or whether trade-offs emerge among them. Within female-anchored configurations, we observe the EER gap decreasing steadily from F+25\%M to F+75\%M (1.940, 1.042, 0.438 pp), while SPD instead increases over the same range (0.0252, 0.0253, 0.0277). We interpret this divergence as reflecting two distinct underlying quantities: the EER gap captures threshold calibration error relative to the dominant gender's acoustic space, whereas SPD captures detection rate disparity across the full score distribution. Improving one is therefore shown to come at the expense of the other, consistent with established fairness impossibility results \cite{chouldechova2017fair}. Under same-attack evaluation, we find that Combined training achieves the best value on six of seven metrics (EER gap = 0.273 pp, FPR gap = 0.0052, EOpD = 0.0042, EOD = 0.0061, PPD = 0.0008, TED = 0.098), while M achieves the lowest SPD (0.0105 versus 0.0108 for Combined). Our results show that the fairness metrics do not move in unison.
\begin{table*}[!t]
\caption{Performance evaluation of six threshold 
calibration strategies in terms of per-gender FPR and 
TPR (\%) across all nine gender-composition 
configurations.}
\label{tab:tc_fpr_tpr}
\centering
\renewcommand{\arraystretch}{1.15}
\setlength{\tabcolsep}{2.5pt}
\small
\begin{tabular*}{\textwidth}{@{\extracolsep{\fill}}
|l|cc|cc|cc|cc|cc|cc|}
\hline
\multirow{3}{*}{Config}
& \multicolumn{2}{c|}{TC0}
& \multicolumn{2}{c|}{TC\_EER}
& \multicolumn{2}{c|}{TC\_FPR}
& \multicolumn{2}{c|}{TC\_TPR}
& \multicolumn{2}{c|}{TC\_DP}
& \multicolumn{2}{c|}{Oracle}\\
\cline{2-13}
& F & M
& F & M
& F & M
& F & M
& F & M
& F & M \\
\hline
\multicolumn{13}{|l|}{\textit{FPR (\%) — lower = fewer false alarms on bonafide speakers}} \\
\hline
F
  & 2.643 & 4.862 & 2.777 & 4.688
  & \textbf{3.753} & \textbf{3.753}
  & 2.069 & 5.791 & 3.227 & 4.361 & 2.777 & 4.688 \\

M
  & 4.454 & 3.688 & 5.052 & 2.778
  & \textbf{4.071} & \textbf{4.071}
  & 6.657 & 2.002 & 4.075 & 4.177 & 5.052 & 2.778 \\

  Combined
  & 2.067 & 2.358 & 2.222 & 2.176
  & \textbf{2.212} & \textbf{2.213}
  & 5.994 & 5.306 & 1.991 & 2.599 & 2.222 & 2.176 \\
  
\hline
F+25\%M
  & 1.172 & 3.703 & 1.253 & 3.410
  & \textbf{1.172} & \textbf{1.173}
  & 0.917 & 4.773 & 5.764 & 3.074 & 1.229 & 3.168 \\
F+50\%M
  & 1.084 & 2.270 & 1.116 & 2.153
  & \textbf{1.084} & \textbf{1.079}
  & 0.905 & 2.826 & 6.199 & 1.776 & 1.145 & 2.201 \\
F+75\%M
  & 1.169 & 1.615 & 1.159 & 1.605
  & \textbf{1.169} & \textbf{1.165}
  & 1.018 & 1.804 & 6.923 & 1.198 & 1.157 & 1.583 \\
M+75\%F
  & 1.828 & 0.920 & 1.668 & 1.037
  & \textbf{1.828} & \textbf{1.823}
  & 1.897 & 0.881 & 7.549 & 0.609 & 1.687 & 1.070 \\
M+50\%F
  & 2.575 & 0.896 & 2.129 & 1.133
  & \textbf{2.575} & \textbf{2.582}
  & 2.902 & 0.810 & 8.040 & 0.613 & 2.339 & 1.122 \\
M+25\%F
  & 4.176 & 0.945 & 3.433 & 1.161
  & \textbf{4.176} & \textbf{4.171}
  & 4.937 & 0.776 & \textit{9.449} & 0.616 & 3.315 & 1.183 \\
\hline
\multicolumn{13}{|l|}{\textit{TPR (\%) — higher = better spoof detection rate}} \\
\hline
F
  & 96.912 & 95.435 & 97.205 & 95.290 & 97.851 & 94.041
  & \textbf{96.178} & \textbf{96.171}
  & 97.056 & 94.830 & 97.205 & 95.290 \\

M
  & 94.237 & 97.760 & 94.926 & 97.205 & 93.835 & 97.997
  & \textbf{96.000} & \textbf{96.000}
  & 93.426 & 97.941 & 94.926 & 97.205 \\
  Combined
  & 97.574 & 97.989 & 97.762 & 97.805 & 97.749 & 97.831
  & \textbf{97.780} & \textbf{97.784}
  & 96.653 & 98.010 & 97.762 & 97.805 \\
\hline
F+25\%M
  & 98.748 & 97.281 & 98.795 & 96.995
  & \textit{92.384} & \textit{84.781}
  & \textbf{98.013} & \textbf{98.015}
  & 99.140 & 96.191 & 98.795 & 96.857 \\
F+50\%M
  & 98.772 & 97.900 & 98.828 & 97.737
  & \textit{89.273} & \textit{85.446}
  & \textbf{98.336} & \textbf{98.337}
  & 99.281 & 96.583 & 98.876 & 97.829 \\
F+75\%M
  & 98.877 & 98.439 & 98.858 & 98.404
  & \textit{92.511} & \textit{90.765}
  & \textbf{98.657} & \textbf{98.659}
  & 99.437 & 96.942 & 98.866 & 98.440 \\
M+75\%F
  & 98.506 & 98.628 & 98.269 & 98.820
  & \textit{92.177} & \textit{93.234}
  & \textbf{98.566} & \textbf{98.568}
  & 99.250 & 96.944 & 98.339 & 98.957 \\
M+50\%F
  & 98.010 & 98.481 & 97.428 & 98.865
  & \textit{97.907} & \textit{99.659}
  & \textbf{98.245} & \textbf{98.245}
  & 98.908 & 96.694 & 97.689 & 98.905 \\
M+25\%F
  & 97.461 & 98.425 & 96.611 & 98.798
  & \textit{97.397} & \textit{99.815}
  & \textbf{97.942} & \textbf{97.943}
  & 98.505 & 96.543 & 96.716 & 98.840 \\
\hline
\end{tabular*}
\end{table*}

\subsubsection{Cross-Attack Generalization}
\label{subsubsec:cross_attack_fairness}

\begin{figure*}[!t]
\centering
\includegraphics[width=\textwidth]{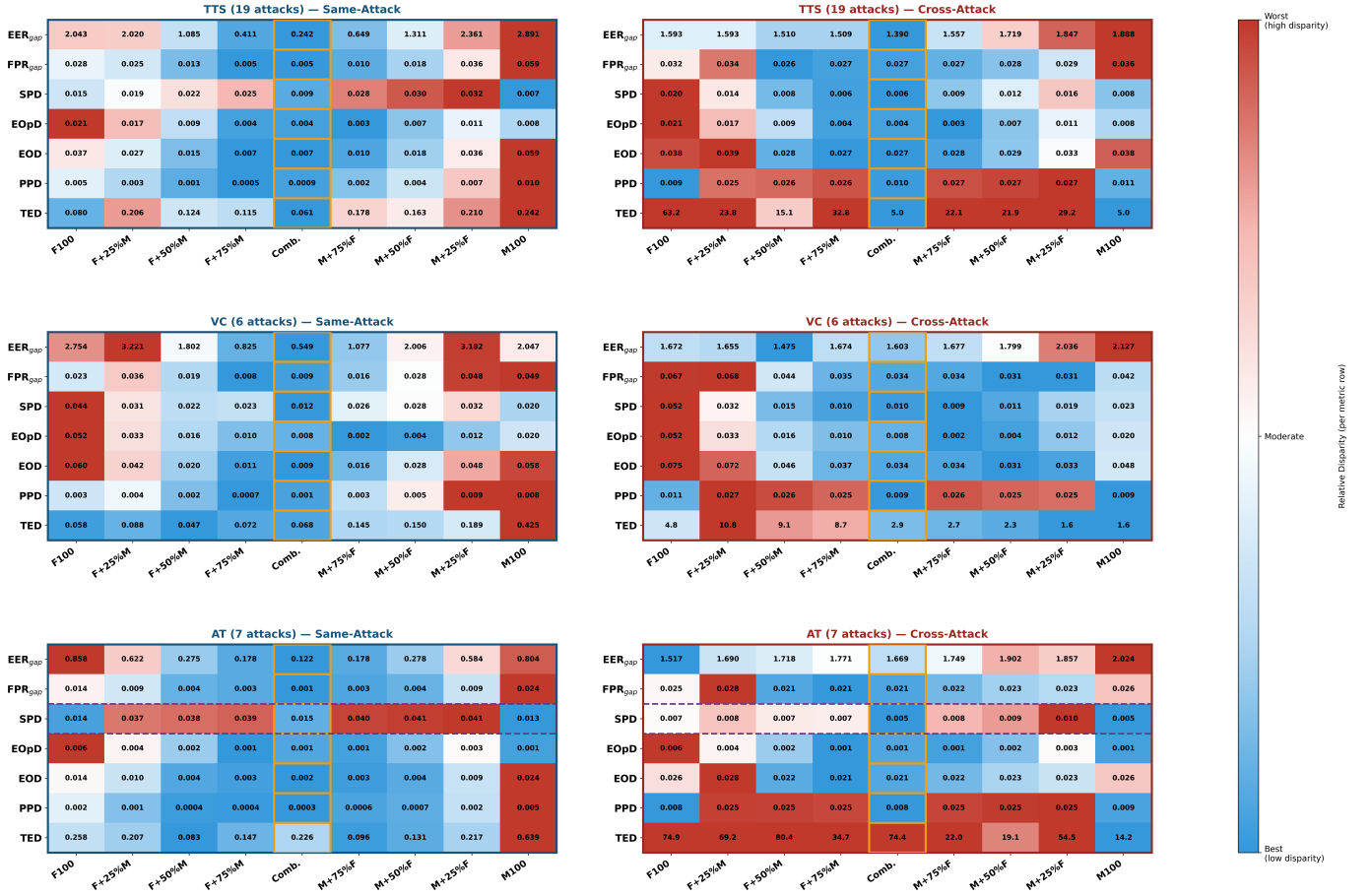}
\caption{Fairness metric heatmap across TTS 
(text-to-speech), VC (voice conversion), and AT 
(adversarial perturbation) attacks under same-attack 
(left side) and cross-attack (right side) 
evaluation across all nine gender-composition 
configurations.}
\label{fig:heatmap_same_vs_cross}
\end{figure*}

Here, we investigate whether trends persist when models are evaluated on unseen attack types. Table~\ref{tab:fairness_cross_all9} and 
Fig.~\ref{fig:heatmap_same_vs_cross} report metrics when 
each model trained on attack $i$ evaluates on all remaining 
attacks $j \neq i$.

Our experimental results reveal that the composition advantages are reduced dramatically under this cross-attack evaluation test. The range of same-attack EER gaps, which spans 0.273–2.276 pp, reduces to 1.491–1.963 pp under cross-attack evaluation.
Furthermore, the ratio between Combined and M EER falls from 8.3x to just 1.3x. We also observe that the monotonic trend anchored by female speakers, which held cleanly under same-attack evaluation, no longer holds in cross-attack evaluation. As we observe that F+50\%M (1.549 pp) now produces a lower bias gap than F+75\%M (1.597 pp), reversing the ordering we reported in Section~\ref{subsubsec:same_attack_fairness}. As can be seen in the table, the male-anchored EER gap pattern remains the same in cross-attack evaluation, though compressed.
Additionally, we observe a more pronounced effect when we examine which gender is disadvantaged. Two of the three female-anchored mixtures switch the disadvantaged gender completely: F+50\%M and F+75\%M disadvantage male speakers under same-attack evaluation but disadvantage female speakers under cross-attack evaluation. In contrast, F+25\%M disadvantages male speakers in both settings. An analysis based only on same-attack evaluation cannot reveal this change, which raises the question of what causes it. We hypothesize that a calibration effect is responsible. Specifically, decision thresholds learned from the score distribution of one attack, which reflects that attack's particular gender bias, may be positioned differently relative to the score distribution of another attack. As a result, the disadvantaged gender can change when the model is evaluated on an unseen attack. AT attacks show 
the sharpest such reversal -- the fairest under same-attack 
evaluation (0.122\,pp) but among the most biased under 
cross-attack evaluation (1.669\,pp) -- because adversarial 
perturbations calibrate tightly to a specific attack's 
feature space and misalign more severely on unseen attacks 
than TTS or VC artifacts, which generalize more gracefully 
across attack types.
Fig. ~\ref{fig:heatmap_same_vs_cross} examines this pattern metric by metric. We find that Combined training attains the lowest or second-lowest value across most metrics in all six panels, suggesting that its advantage generalizes reasonably well, although it becomes less pronounced under cross-attack evaluation.We further observe that AT SPD effectively vanishes under cross-attack evaluation, surviving only as statistical ties in Table~\ref{tab:fairness_cross_all9}; this confirms our earlier suspicion that the SPD inversion reported in Table~\ref{tab:fairness_all9} was an artifact of within-attack threshold calibration rather than a genuine fairness effect. VC attacks, by contrast, continue to dominate most metrics under cross-attack evaluation and retain comparatively high FPR gap values in both settings. As shown in Fig.~\ref{fig:heatmap_same_vs_cross}, TED scores increase sharply across all configurations compared with their same-attack values: F rises from 0.115 to 54.785, Combined from 0.098 to 19.819, and M from 0.363 to 6.358. We attribute this increase to numerical instability in the underlying FP/FN ratio measured by TED. Specifically, near-zero false-negative counts at operating points close to the cross-attack EER threshold cause large variations in the metric. Therefore, we interpret cross-attack TED values with caution and exclude them from the subsequent cross-attack comparisons.

\subsubsection{Attack-Level Bias Patterns}
\label{subsubsec:attack_patterns}
Here, we investigate attack-level bias patterns by examining whether all spoofing attacks are equally affected. Table~\ref{tab:fair_unfair_1} lists all 32 attacks in rank order by their mean rank across SPD, EOpD, EOD, and PPD, and ranks TED separately. We find that four of the five most biased attacks according to the four-metric ranking are pre-trained TTS systems. A28 (pre-trained YourTTS) ranks as the most biased attack overall, both by mean rank across the four metrics and individually, with EOD = 0.105 and an EER gap of 6.759 pp. We also find that A17 (ZMM-TTS) and A29 (XTTS) rank among the five most biased attacks, which we attribute to their similar pre-training process \cite{c21}. In addition, A13 (StarGANv2-VC) appears in this group despite not relying on pre-training; we hypothesize that this behavior reflects the gender-based score distribution shift that prior work has linked to domain shift in voice conversion systems \cite{c21}.

We further observe that ranking the same 32 attacks by TED produces a markedly different scenario: none of the attacks in the TED-worst list appears in the four-metric-worst list. For example, A19 (MaryTTS) ranks 32nd by TED yet ranks as the ninth fairest attack according to the four rate-based metrics, while A14 (from-scratch YourTTS) ranks as the second fairest attack by error-rate metrics but only 29th by TED. Specifically, comparing A28 and A14 keeps the underlying framework fixed, as both attacks use YourTTS, and changes only whether the system was pre-trained. The difference is substantial: A28 produces an EER gap of 6.759 pp, whereas A14 produces only 0.031 pp, a 218-fold difference attributable solely to pre-training. Together with our findings for the WavLM-Base+ baseline, this comparison suggests that pre-training is a major source of unintentional gender-correlated information \cite{c19, hutiri2022bias}. In TTS synthesis, pre-training creates attacks with gendered prosody, whereas in self-supervised representations, it enables the detector to learn gender-correlated cues. We conclude that the choice of fairness metric strongly influences which attacks appear biased, highlighting that no single metric is sufficient to characterize fairness at the attack level.

\begin{table}[!t]
\caption{Ranking comparison of attacks by fairness metrics 
along with EER gap. Top and bottom panels rank attacks by 
mean rank across SPD, EOpD, EOD, and PPD; middle panel 
ranks attacks by TED separately.}
\label{tab:fair_unfair_1}
\centering
\renewcommand{\arraystretch}{1.15}
\setlength{\tabcolsep}{2.5pt}
\small
\resizebox{\columnwidth}{!}{%
\begin{tabular}{|c|c|c|c|c|c|c|}
\hline
Attack &Type & SPD 
& EOpD & EOD & PPD 
& EER gap$^{\dagger}$ \\
\hline
\multicolumn{7}{l}{\textit{Most Unfair --- by SPD, EOpD, EOD, and PPD}} \\
\hline
A28 & TTS & 0.038 & 0.034 & 0.105 & 0.020 & 6.759\,pp \\
A13 & VC  & 0.035 & 0.045 & 0.063 & 0.006 & 3.789\,pp \\
A17 & TTS & 0.027 & 0.017 & 0.063 & 0.012 & 3.455\,pp \\
A15 & VC  & 0.033 & 0.039 & 0.058 & 0.009 & 3.489\,pp \\
A29 & TTS & 0.024 & 0.014 & 0.047 & 0.008 & 2.899\,pp \\
\hline
\multicolumn{7}{|l|}{\textit{Most Unfair --- by TED only}} \\
\hline
A19 & TTS & \multicolumn{4}{c|}{TED = 0.402} & 0.355\,pp \\
A32 & AT  & \multicolumn{4}{c|}{TED = 0.313} & 0.098\,pp \\
A30 & AT  & \multicolumn{4}{c|}{TED = 0.278} & 0.588\,pp \\
A14 & VC  & \multicolumn{4}{c|}{TED = 0.258} & 0.031\,pp \\
A25 & VC  & \multicolumn{4}{c|}{TED = 0.248} & 1.298\,pp \\
\hline
\multicolumn{7}{|l|}{\textit{Most Fair --- by SPD, EOpD, EOD, and PPD}} \\
\hline
A14 & VC  & 0.013 & 0.0002 & 0.001 & 0.0001 & 0.031\,pp \\
A12 & TTS & 0.013 & 0.0010 & 0.002 & 0.0002 & 0.114\,pp \\
A16 & VC  & 0.013 & 0.0050 & 0.007 & 0.0010 & 0.422\,pp \\
A20 & AT  & 0.014 & 0.0020 & 0.004 & 0.0010 & 0.143\,pp \\
A19 & TTS & 0.008 & 0.0020 & 0.003 & 0.0010 & 0.355\,pp \\
\hline
\multicolumn{7}{|l|}{$^{\dagger}$EER gap is shown for reference only 
and is not used in the ranking.}
\end{tabular}%
}
\end{table}

%=============================================================
\subsection{Mitigation Strategies : Threshold Calibration Results}
\label{subsec:tc_results}
%=============================================================
We next explore whether post-hoc threshold calibration can close the gender fairness gap without retraining. We observed two consistent findings. First, the EER gap is completely resistant to calibration. As Table~\ref{tab:tc_fairness} shows, the EER gap remains fixed at 1.317 pp across all six strategies, including Oracle calibration with full access to evaluation-set labels. This result confirms and extends the theoretical findings of \cite{hardt2016equality}, who showed that post-hoc calibration can satisfy individual fairness criteria but cannot eliminate structural disparities in the underlying score distributions.

Second, we find that each calibration strategy succeeds in its intended goal, but this success comes with trade-offs in other metrics. As can be seen, TC\_FPR effectively equalizes false alarms and reduces FPR\_gap to nearly zero (0.0001), but this improvement comes at a clear cost to detection performance. Table~\ref{tab:tc_fairness} shows that TC\_FPR reduces TPR to 84- 92\% under GenderMix configurations, compared with 97 -99\% under TC0. We interpret this behavior as a direct consequence of increasing the threshold to equalize false alarms, which necessarily reduces both genders' ability to detect spoofed speech. Surprisingly, TC\_TPR exhibits the opposite trade-off. By equalizing spoof detection rates across genders, it reduces EOpD to 0.0001 but increases FPR\_gap to 0.0261, compared with 0.0152 under TC0. We consider TC\_TPR to be most appropriate for deployments in which missed spoof detections are more critical than false alarms.

Table~\ref{tab:tc_fpr_tpr} shows that the right calibration choice depends on which mistake hurts more: a false alarm on a real speaker, or a missed spoof attack. TC\_FPR keeps false alarms equal across genders in every configuration, such as 3.753\% for both under F, which suits something like consumer voice login. The cost lands on detection: under female-skewed configurations, spoof detection drops to 84\% to 92\%, down from 97\% to 99\% without calibration. TC\_TPR does the opposite, keeping detection rates nearly identical across genders, within a tenth of a point, which fits higher-stakes settings like financial or forensic verification. Its cost falls on false alarms instead, which can differ by as little as under 1 point or as much as 4.66 points depending on the configuration. TC\_DP fits neither story well: it pushes male false alarms near zero (0.616\%) while female false alarms climb past 9\%, making it a poor fit regardless of which mistake matters more.

Finally, TC\_EER reveals an unexpected dependency. Under balanced training, our previous work \cite{fursule2026smc} showed that TC\_EER reduces disparity by 54-75\%. Under the varied compositions examined here, however, TC\_EER provides only a marginal improvement in FPR\_gap (0.0137 versus 0.0152 under TC0) and actually worsens SPD (0.0276 versus 0.0237). We attribute this failure to the composition-dependent score distribution shifts discussed in Section~\ref{subsec:composition} and conclude that TC\_EER should be applied only under combined, balanced training rather than under the imbalanced configurations where it appears most needed. Lastly, we investigate whether thresholds calibrated for one attack generalize to others. We calibrate development-set thresholds on the same attack used for training and apply them unchanged to unseen attack types. Under this mismatch, FPR rises above 70\%, indicating that threshold calibration depends on attack-specific score distributions and requires recalibration whenever the deployment attack changes.

\subsubsection{Practical Guidelines for Trustworthy 
Deployment}
\label{subsec:disc_deployment}

Table~\ref{tab:metric_corr} shows that SPD is 
statistically independent of EER gap ($\rho = +0.065$, 
$p = 0.270$) and of EOD, PPD, and FPR\_gap, while EOpD, 
EOD, PPD, and FPR\_gap form a strongly correlated cluster 
($\rho > 0.41$, EOD vs.\ FPR\_gap reaching $\rho = 0.953$). 
TED negatively correlates with EOpD ($\rho = -0.166$, 
$p = 0.005$), indicating that equalizing detection rates 
tends to increase FP/FN asymmetry. Combined with our 
finding in Section~\ref{subsubsec:attack_patterns} that 
metric choice changes the attack-level fairness verdict 
for 5 of 32 attacks~\cite{c22}, we conclude that 
practitioners should select the metric that best matches 
their deployment context. For voice authentication, where 
false rejection equity is the primary concern, we recommend 
prioritizing FPR\_gap, as A28 consistently ranks as the 
worst attack under this criterion. For content moderation, 
where missed spoof detections represent the critical 
failure mode, we recommend prioritizing EOpD, as A13 ranks 
worst. For broadcast monitoring, where the demographic 
reach of flagging decisions must be equitable, we recommend 
prioritizing SPD, as A27 ranks worst under this criterion.

These findings further motivate four minimum requirements 
we propose for trustworthy system certification. First, 
we recommend that practitioners supplement aggregate EER 
with per-gender EER reported separately for each attack 
in the target deployment environment, since aggregate 
metrics can conceal systematic per-gender failures. 
Second, we recommend reporting all six fairness metrics -- 
SPD, EOpD, EOD, PPD, TED, and FPR\_gap -- rather than 
relying on any single one, given that metric choice alone 
can flip the attack-level fairness verdict. Third, we 
recommend documenting training gender composition as 
standard metadata alongside performance numbers, since 
our results in Section~\ref{subsubsec:same_attack_fairness} 
show that composition is the strongest predictor of bias 
direction. Fourth, we recommend deliberate calibration 
strategy selection: practitioners should apply TC\_FPR 
when false alarm equity is the deployment priority, 
TC\_TPR when detection rate equity matters most, and 
avoid TC\_EER entirely under imbalanced training 
conditions.

Two practical constraints bound these guidelines. Threshold calibration requires speaker gender labels at inference time, which many deployments do not provide, and the binary gender framing used here reflects the ASVspoof5 protocol labels and does not extend to non-binary or intersectional identities. Within these bounds, our results map which fairness gaps training composition predicts, which feature choice determines, and which post-hoc strategies can and cannot close

 %=======================================
\begin{table}[!t]
\centering
\caption{Performance comparison of fairness metrics across six threshold calibration strategies, averaged over all nine gender-composition configurations and 32 
same-attack evaluations. }
\label{tab:tc_fairness}
\renewcommand{\arraystretch}{1.15}
\setlength{\tabcolsep}{3pt}
\scriptsize
\resizebox{\columnwidth}{!}{%
\begin{tabular}{|l|c|c|c|c|c|c|}
\hline
Strategy
& EER gap
& FPR\_gap
& SPD
& EOpD
& EOD
& PPD \\
\hline
TC0     & 1.317 & 0.0152 & 0.0237 & 0.0124 & 0.0203 & 0.0031 \\
TC\_EER & 1.317 & 0.0137 & 0.0276 & 0.0136 & 0.0162 & 0.0024 \\
TC\_FPR & 1.317 & \textbf{0.0001} & 0.0293 & 0.0301 & 0.0301 & \textbf{0.0021} \\
TC\_TPR & 1.317 & 0.0261 & 0.0304 & \textbf{0.0001} & 0.0261 & 0.0027 \\
TC\_DP  & 1.317 & 0.0479 & \textbf{0.0000} & 0.0268 & 0.0573 & 0.0099 \\
Oracle  & 1.317 & 0.0132 & 0.0268 & 0.0132 & \textbf{0.0133} & 0.0023 \\
\hline
\end{tabular}%
}
\end{table}

\begin{table}[!t]
\centering
\caption{Pairwise Pearson correlations between fairness metrics and EER gap across all 288 WavLM same-attack evaluations. $^{*}p < 0.05$; $^{**}p < 0.001$; ns\,=\,not significant.}
\label{tab:metric_corr}
\renewcommand{\arraystretch}{1.15}
\setlength{\tabcolsep}{4pt}
\small
\begin{tabular}{|l|c|c|}
\hline
Metric Pair & \textbf{$\rho$} & \textbf{$p$} \\
\hline
\multicolumn{3}{|l|}{\textit{SPD — independent of all others}} \\
\hline
SPD vs.\ EER gap    & $+0.065$ & $0.270$\,ns \\
SPD vs.\ EOpD        & $+0.376$ & $<0.001^{**}$ \\
SPD vs.\ EOD         & $+0.083$ & $0.160$\,ns \\
SPD vs.\ FPR\_gap    & $-0.063$ & $0.287$\,ns \\
\hline
\multicolumn{3}{|l|}{\textit{EOpD/EOD/PPD/FPR\_gap — strongly correlated cluster}} \\
\hline
EOD vs.\ FPR\_gap    & $+0.953$ & $<0.001^{**}$ \\
PPD vs.\ FPR\_gap    & $+0.974$ & $<0.001^{**}$ \\
EER gap vs.\ EOD    & $+0.932$ & $<0.001^{**}$ \\
EER gap vs.\ FPR\_gap & $+0.922$ & $<0.001^{**}$ \\
\hline
\multicolumn{3}{|l|}{\textit{TED — orthogonal to rate-based metrics}} \\
\hline
TED vs.\ EOpD        & $-0.166$ & $0.005^{*}$ \\
TED vs.\ EER gap    & $+0.056$ & $0.340$\,ns \\
TED vs.\ FPR\_gap    & $+0.230$ & $<0.001^{**}$ \\
\hline
\end{tabular}
\end{table}

% ============================================================
\section{Limitations}
\label{sec:limitations}
% =============

In this section, we discuss the limitations of this study. Five limitations define the scope of our findings. 

First, we rely on the binary gender labels provided by 
the official ASVspoof5 protocol files, which restricts 
our analysis to female and male speaker subgroups and 
excludes non-binary, self-identified, and intersectional 
gender identities. 

Second, we construct custom 70/10/20 splits that provide 
a controlled setup for fairness analysis but are not 
directly comparable to official ASVspoof5 challenge 
scores. We stratify splits by attack and gender but do 
not guarantee speaker disjointness across splits, potential speaker overlap across splits is acknowledged as a limitation in Section~\ref{sec:limitations}. We emphasize that the primary goal of this study is not speaker-independent generalization, but controlled comparison of gender-composition effects under identical split construction.. Because the evaluation pool constitutes 20\% of all available 
utterances and ASVspoof5 contains multiple utterances per 
speaker, we expect evaluation speakers to also appear in 
the training pool. Speaker IDs are not available in the 
evaluation outputs we use here, so we cannot quantify the 
exact fraction of overlapping speakers; the direction and 
magnitude of any resulting bias in our fairness estimates 
therefore remain unknown.

Third, we compare only two feature representations: 
LogSpectrogram, evaluated under the three baseline 
configurations, and WavLM-Base+, evaluated under all 
nine configurations. Whether the composition-driven 
fairness patterns we observe generalize to other 
self-supervised representations or model architectures 
remains an open question we leave to future work.

Fourth, we calibrate threshold calibration strategies on 
same-attack development data and apply them to evaluation 
utterances from that same attack. We do not evaluate how 
well these thresholds generalize to unseen attack types 
at deployment time, and we identify attack-matched 
cross-attack calibration as an important direction for 
future work.

Fifth, we evaluate post-hoc mitigation only. Calibration 
adjusts decision thresholds but cannot alter learned 
representations, training objectives, or score 
distributions. Training-time interventions such as 
fairness-aware objectives, adversarial debiasing, and 
representation-level regularization fall outside the 
scope of this study and represent the most pressing 
direction for future work, given our finding that 
threshold calibration cannot close distribution-level 
EER gaps regardless of how well the calibration strategy 
is designed.
 
% ============================================================
\section{Conclusion}
\label{sec:conclusion}
This paper presents four main findings. First, training data composition strongly predicts bias direction: the gender underrepresented in training consistently performs worse at test time, regardless of which gender occupies the minority position. Second, feature representation shapes bias magnitude independently of training composition: WavLM-Base+ produces gender gaps 3.0 to 4.3 times larger than LogSpectrogram under identical training conditions, and balanced training reduces LogSpectrogram bias but leaves WavLM bias largely intact. Third, post-hoc threshold calibration cannot close the EER gap: all six strategies tested, including Oracle calibration with full test-set label access, leave the EER gap unchanged at 1.317\,pp, confirming that distribution-level disparities require retraining rather than recalibration. Fourth, no single fairness metric captures every failure mode: metric choice alone flips the attack-level fairness verdict for 5 of 32 attacks, and per-gender EER calibration that works well under balanced training actively worsens equalized odds under imbalanced training. Together, these findings indicate that gender fairness in audio deepfake detection must be built in at training time, as post-hoc correction can only partially address the resulting disparities.


\begin{thebibliography}{1}

\bibitem{mai2023humans}
K.~T.~Mai, S.~Bray, T.~Davies, and L.~D.~Griffin,
``Warning: Humans cannot reliably detect speech deepfakes,''
\emph{PLOS ONE},
vol.~18, no.~8, p.~e0285333, 2023.

\bibitem{c1}
O.~A. Shaaban and R.~Yildirim,
``Audio deepfake detection using deep learning,''
\emph{Engineering Reports}, vol.~7, no.~3, p.~e70087, 2025.

\bibitem{c2}
I.~Khan, K.~Khan, and A.~Ahmad,
``A comprehensive survey of deepfake generation and detection techniques in audio-visual media,''
\emph{ICCK Journal of Image Analysis and Processing}, vol.~1, no.~2, pp.~73--95, 2025.

\bibitem{yi2023audio}
J.~Yi, C.~Wang, J.~Tao, X.~Zhang, C.~Y.~Zhang, and Y.~Zhao,
``Audio deepfake detection: A survey,''
\emph{arXiv preprint arXiv:2308.14970}, 2023.

\bibitem{hong2026vulnerabilities}
M.~Hong, D.~Jiang, Z.~Xie, W.~Zhao, G.~Wang, and C.~J.~Zhang,
``Vulnerabilities of audio-based biometric authentication systems against
deepfake speech synthesis,''
\emph{arXiv preprint arXiv:2601.02914}, 2026.

\bibitem{zhang2025audio}
B.~Zhang, H.~Cui, V.~Nguyen, and M.~Whitty,
``Audio deepfake detection: What has been achieved and what lies ahead,''
\emph{Sensors},
vol.~25, no.~7, p.~1989, 2025.


\bibitem{c4}
X.~Wang, H.~Delgado, H.~Tak, J.~W. Jung, H.~J. Shim, M.~Todisco, et~al.,
``ASVspoof 5: Crowdsourced speech data, deepfakes, and adversarial attacks at scale,''
\emph{arXiv preprint arXiv:2408.08739}, 2024.

\bibitem{c6}
H.~Tak, J.~Patino, M.~Todisco, A.~Nautsch, N.~Evans, and A.~Larcher,
``End-to-end anti-spoofing with RawNet2,''
in \emph{Proc. ICASSP 2021 -- IEEE International Conference on Acoustics, Speech and Signal Processing (ICASSP)},
2021, pp.~6369--6373.


\bibitem{c7}
J.~W. Jung, H.~S. Heo, H.~Tak, H.~J. Shim, J.~S. Chung, B.~J. Lee, et~al.,
``AASIST: Audio anti-spoofing using integrated spectro-temporal graph attention networks,''
in \emph{Proc. ICASSP 2022 -- IEEE International Conference on Acoustics, Speech and Signal Processing (ICASSP)},
2022, pp.~6367--6371.

\bibitem{c13}
E.~Ntoutsi, P.~Fafalios, U.~Gadiraju, V.~Iosifidis, W.~Nejdl, M.~E. Vidal, et~al.,
``Bias in data-driven artificial intelligence systems: An introductory survey,''
\emph{Wiley Interdisciplinary Reviews: Data Mining and Knowledge Discovery},
vol.~10, no.~3, p.~e1356, 2020.

\bibitem{c14}
G.~Fenu, M.~Marras, G.~Medda, and G.~Meloni,
``Fair voice biometrics: Impact of demographic imbalance on group fairness in speaker recognition,''
in \emph{Proc. Interspeech},
2021, pp.~1892--1896.

\bibitem{c15}
J.~J. Bird and A.~Lotfi,
``Real-time detection of AI-generated speech for deepfake voice conversion,''
\emph{arXiv preprint arXiv:2308.12734}, 2023.

\bibitem{c18}
A.~Fursule, S.~Kshirsagar, and A.~R. Avila,
``Gender fairness in audio deepfake detection: Performance and disparity analysis,''
in \emph{Proc. 2026 IEEE Conference on Artificial Intelligence (CAI)},
2026, pp.~2116--2121.

\bibitem{c19}
T.~Hashimoto, M.~Srivastava, H.~Namkoong, and P.~Liang,
``Fairness without demographics in repeated loss minimization,''
in \emph{Proc. International Conference on Machine Learning},
2018, pp.~1929--1938.

\bibitem{c21}
H.~S. Nguyen-Le, H.~C. Nguyen-Thanh, N.~A. Le-Khac, D.~T. Nguyen, and H.~H. Nguyen-Le,
``AFSS: Artifact-focused self-synthesis for mitigating bias in audio deepfake detection,''
\emph{arXiv preprint arXiv:2603.26856}, 2026.

\bibitem{c25}
A.~V. Nadimpalli and A.~Rattani,
``GBDF: Gender balanced deepfake dataset towards fair deepfake detection,''
in \emph{Proc. International Conference on Pattern Recognition},
2022, pp.~320--337.

\bibitem{c22}
N.~Mehrabi, F.~Morstatter, N.~Saxena, K.~Lerman, and A.~Galstyan,
``A survey on bias and fairness in machine learning,''
\emph{ACM Computing Surveys}, vol.~54, no.~6, pp.~1--35, 2021. 

\bibitem{c26}
Y.~Ju, S.~Hu, S.~Jia, G.~H. Chen, and S.~Lyu,
``Improving fairness in deepfake detection,''
in \emph{Proc. IEEE/CVF Winter Conference on Applications of Computer Vision},
2024, pp.~4655--4665.

\bibitem{c27}
L.~Lin, X.~He, Y.~Ju, X.~Wang, F.~Ding, and S.~Hu,
``Preserving fairness generalization in deepfake detection,''
in \emph{Proc. IEEE/CVF Conference on Computer Vision and Pattern Recognition},
2024, pp.~16815--16825.

\bibitem{yadav2024fairssd}
A.~K.~S. Yadav, K.~Bhagtani, D.~Salvi, P.~Bestagini, and E.~J. Delp,
``FairSSD: Understanding bias in synthetic speech detectors,''
in \emph{Proc. IEEE/CVF Conference on Computer Vision and Pattern Recognition (CVPR)},
2024, pp.~4418--4428.

\bibitem{hardt2016equality}
M.~Hardt, E.~Price, and N.~Srebro,
``Equality of opportunity in supervised learning,''
in \emph{Advances in Neural Information Processing Systems},
vol.~29, 2016.

\bibitem{chouldechova2017fair}
A.~Chouldechova,
``Fair prediction with disparate impact: A study of bias in recidivism prediction instruments,''
\emph{Big Data}, vol.~5, no.~2, pp.~153--163, 2017.

\bibitem{c29}
D.~E. Temmar, A.~Hamadene, V.~Nallaguntla, A.~Fursule, M.~S. Allili, S.~Kshirsagar, and A.~R. Avila,
``Phonetic analysis of real and synthetic speech using HuBERT embeddings: Perspectives for deepfake detection,''
in \emph{Proc. 2025 IEEE International Conference on Systems, Man, and Cybernetics (SMC)},
2025, pp.~86--91.

\bibitem{fursule2026smc}
A.~Fursule, S.~Kshirsagar, and A.~R. Avila,
``Towards trustworthy audio deepfake detection:
A systematic framework for diagnosing and mitigating
gender bias,''
in \emph{Proc. IEEE International Conference on
Systems, Man, and Cybernetics (SMC)}, 2026.

\bibitem{c30}
V.~Nallaguntla, A.~Fursule, S.~Kshirsagar, and A.~R. Avila,
``PhonemeDF: A synthetic speech dataset for audio deepfake detection and naturalness evaluation,''
\emph{arXiv preprint arXiv:2603.15037}, 2026.

\bibitem{c31}
S.~Kshirsagar and A.~R. Avila,
``Investigating the impact of speech enhancement on audio deepfake detection in noisy environments,''
\emph{arXiv preprint arXiv:2603.14767}, 2026.

\bibitem{trinh2021examination}
L.~Trinh and Y.~Liu,
``An examination of fairness of AI models for deepfake detection,''
\emph{arXiv preprint arXiv:2105.00558}, 2021.

\bibitem{xu2024analyzing}
Y.~Xu, P.~Terh{\"o}rst, M.~Pedersen, and K.~Raja,
``Analyzing fairness in deepfake detection with massively annotated databases,''
\emph{IEEE Transactions on Technology and Society}, vol.~5, no.~1, pp.~93--106, 2024.

\bibitem{hutiri2022bias}
W.~T. Hutiri and A.~Y. Ding,
``Bias in automated speaker recognition,''
in \emph{Proc. ACM Conference on Fairness, Accountability, and Transparency (FAccT)},
2022, pp.~230--247.


\bibitem{stanvek2025scdf} V.~Stan{\v{e}}k, K.~Srna, A.~Firc, and K.~Malinka,
``SCDF: A Speaker Characteristics DeepFake Speech Dataset for Bias Analysis,''
\emph{arXiv preprint arXiv:2508.07944}, 2025.

\bibitem{he2016resnet}
K.~He, X.~Zhang, S.~Ren, and J.~Sun,
``Deep residual learning for image recognition,''
in \emph{Proc. IEEE Conference on Computer Vision and Pattern Recognition},
2016, pp.~770--778.

\bibitem{pascu2025easy}
O.~Pascu, D.~Onea{\c{t}}{\u{a}}, H.~Cucu, and N.~M{\"u}ller,
``Easy, interpretable, effective: openSMILE for voice deepfake detection,''
in \emph{Proc. ICASSP 2025 -- IEEE International Conference on Acoustics, Speech and Signal Processing (ICASSP)},
2025, pp.~1--5.

\bibitem{tan2024naturalspeech}
X.~Tan, J.~Chen, H.~Liu, J.~Cong, C.~Zhang, Y.~Liu, et~al.,
``NaturalSpeech: End-to-end text-to-speech synthesis with human-level quality,''
\emph{IEEE Transactions on Pattern Analysis and Machine Intelligence},
vol.~46, no.~6, pp.~4234--4245, 2024.

\bibitem{wu2017asvspoof}
Z.~Wu, J.~Yamagishi, T.~Kinnunen, C.~Hanil{\c{c}}i, M.~Sahidullah, A.~Sizov, et~al.,
``ASVspoof: The automatic speaker verification spoofing and countermeasures challenge,''
\emph{IEEE Journal of Selected Topics in Signal Processing},
vol.~11, no.~4, pp.~588--604, 2017.

\bibitem{yamagishi2021asvspoof}
J.~Yamagishi, X.~Wang, M.~Todisco, M.~Sahidullah, J.~Patino, A.~Nautsch, et~al.,
``ASVspoof 2021: Accelerating progress in spoofed and deepfake speech detection,''
\emph{arXiv preprint arXiv:2109.00537}, 2021.

\bibitem{liu2023asvspoof}
X.~Liu, X.~Wang, M.~Sahidullah, J.~Patino, H.~Delgado, T.~Kinnunen, et~al.,
``ASVspoof 2021: Towards spoofed and deepfake speech detection in the wild,''
\emph{IEEE/ACM Transactions on Audio, Speech, and Language Processing},
vol.~31, pp.~2507--2522, 2023.

\bibitem{peri2022adversarial}
R.~Peri, K.~Somandepalli, and S.~Narayanan,
``To train or not to train adversarially: A study of bias mitigation strategies for speaker recognition,''
\emph{arXiv preprint arXiv:2203.09122}, 2022.

\bibitem{chen2022wavlm}
S.~Chen, C.~Wang, Z.~Chen, Y.~Wu, S.~Liu, Z.~Chen, et~al.,
``WavLM: Large-scale self-supervised pre-training for full stack speech processing,''
\emph{IEEE Journal of Selected Topics in Signal Processing},
vol.~16, no.~6, pp.~1505--1518, 2022.

\bibitem{benjamini1995fdr}
Y.~Benjamini and Y.~Hochberg,
``Controlling the false discovery rate: A practical and powerful approach to multiple testing,''
\emph{Journal of the Royal Statistical Society: Series B (Methodological)},
vol.~57, no.~1, pp.~289--300, 1995.

\bibitem{rubio2026odyssey}
S.~Rubio, P.~Bello, D.~Ribas, A.~Miguel, E.~Lleida, and A.~Ortega,
``An intervention-based framework for shortcut diagnosis in spoofing countermeasures,''
in \emph{Proc. Odyssey 2026: The Speaker and Language Recognition Workshop},
2026, pp.~1--8.

\bibitem{das2026odyssey}
A.~Das, Y.~El~Kheir, F.~R. Guttierez, T.~Polzehl, and S.~M{\"o}ller,
``Can SSL frontend generalize to all-type audio spoofing?''
in \emph{Proc. Odyssey 2026: The Speaker and Language Recognition Workshop}, 2026, pp.~277--283.

\bibitem{loshchilov2019adamw}
I.~Loshchilov and F.~Hutter,
``Decoupled weight decay regularization,''
\emph{arXiv preprint arXiv:1711.05101}, 2017.

\bibitem{hanley2006bootstrap}
J.~A. Hanley and B.~MacGibbon,
``Creating non-parametric bootstrap samples using Poisson frequencies,''
\emph{Computer Methods and Programs in Biomedicine},
vol.~83, no.~1, pp.~57--62, 2006.

\bibitem{benjamini2001fdr}
Y.~Benjamini and D.~Yekutieli,
``The control of the false discovery rate in multiple testing under dependency,''
\emph{The Annals of Statistics},
vol.~29, no.~4, pp.~1165--1188, 2001.

\bibitem{efron1994introduction}
B.~Efron and R.~J. Tibshirani,
\emph{An Introduction to the Bootstrap}.
New York, NY, USA: Chapman and Hall, 1993.

\bibitem{steiner2018marytts}
I.~Steiner and S.~Le Maguer,
``Creating new language and voice components for the updated MaryTTS text-to-speech synthesis platform,''
in \emph{Proceedings of the Eleventh International Conference on Language Resources and Evaluation (LREC 2018)},
Miyazaki, Japan, May 2018.

\bibitem{gong2024zmmtts}
C.~Gong, X.~Wang, E.~Cooper, D.~Wells, L.~Wang, J.~Dang,
and J.~Yamagishi,
``ZMM-TTS: Zero-shot multilingual and multispeaker speech synthesis
conditioned on self-supervised discrete speech representations,''
\emph{IEEE/ACM Transactions on Audio, Speech, and Language Processing},
vol.~32, pp.~4036--4051, 2024.

\bibitem{casanova2022yourtts}
E.~Casanova, J.~Weber, C.~D.~Shulby, A.~C.~Junior, E.~Gölge,
and M.~A.~Ponti,
``YourTTS: Towards zero-shot multi-speaker TTS and zero-shot voice
conversion for everyone,''
in \emph{Proceedings of the International Conference on Machine Learning
(ICML)},
pp.~2709--2720, PMLR, Jun. 2022.

\bibitem{casanova2024xtts}
E.~Casanova, K.~Davis, E.~Gölge, G.~Göknar, I.~Gulea, L.~Hart,
and J.~Weber,
``XTTS: A massively multilingual zero-shot text-to-speech model,''
\emph{arXiv preprint arXiv:2406.04904}, 2024.

\bibitem{kim2020glowtts}
J.~Kim, S.~Kim, J.~Kong, and S.~Yoon,
``Glow-TTS: A generative flow for text-to-speech via monotonic alignment
search,''
in \emph{Advances in Neural Information Processing Systems},
vol.~33, pp.~8067--8077, 2020.

\bibitem{popov2021gradtts}
V.~Popov, I.~Vovk, V.~Gogoryan, T.~Sadekova, and M.~Kudinov,
``Grad-TTS: A diffusion probabilistic model for text-to-speech,''
in \emph{Proceedings of the International Conference on Machine Learning
(ICML)},
pp.~8599--8608, PMLR, Jul. 2021.

\bibitem{lee2022bigvgan}
S.~G.~Lee, W.~Ping, B.~Ginsburg, B.~Catanzaro, and S.~Yoon,
``BigVGAN: A universal neural vocoder with large-scale training,''
\emph{arXiv preprint arXiv:2206.04658}, 2022.

\bibitem{lux2023prosody}
F.~Lux, J.~Koch, and N.~T.~Vu,
``Exact prosody cloning in zero-shot multispeaker text-to-speech,''
in \emph{Proceedings of the 2022 IEEE Spoken Language Technology Workshop
(SLT)},
pp.~962--969, IEEE, Jan. 2023.

\bibitem{lancucki2021fastpitch}
A.~{\L}a{\'n}cucki,
``FastPitch: Parallel text-to-speech with pitch prediction,''
in \emph{Proceedings of the IEEE International Conference on Acoustics,
Speech and Signal Processing (ICASSP)},
pp.~6588--6592, IEEE, Jun. 2021.

\bibitem{kim2021vits}
J.~Kim, J.~Kong, and J.~Son,
``Conditional variational autoencoder with adversarial learning for
end-to-end text-to-speech,''
in \emph{Proceedings of the International Conference on Machine Learning
(ICML)},
pp.~5530--5540, PMLR, Jul. 2021.

\bibitem{lux2022lowresource}
F.~Lux, J.~Koch, and N.~T.~Vu,
``Low-resource multilingual and zero-shot multispeaker TTS,''
in \emph{Proceedings of the 2nd Conference of the Asia-Pacific Chapter of
the Association for Computational Linguistics and the 12th International
Joint Conference on Natural Language Processing (Volume 1: Long Papers)},
pp.~741--751, Nov. 2022.

\bibitem{popov2021diffvc}
V.~Popov, I.~Vovk, V.~Gogoryan, T.~Sadekova, M.~Kudinov,
and J.~Wei,
``Diffusion-based voice conversion with fast maximum likelihood sampling
scheme,''
\emph{arXiv preprint arXiv:2109.13821}, 2021.

\bibitem{kong2020hifigan}
J.~Kong, J.~Kim, and J.~Bae,
``HiFi-GAN: Generative adversarial networks for efficient and high-fidelity
speech synthesis,''
in \emph{Advances in Neural Information Processing Systems},
vol.~33, pp.~17022--17033, 2020.

\bibitem{shen2018tacotron2}
J.~Shen, R.~Pang, R.~J.~Weiss, M.~Schuster, N.~Jaitly, Z.~Yang,
Z.~Chen, Y.~Zhang, Y.~Wang, R.~Skerry-Ryan, R.~A.~Saurous,
Y.~Agiomyrgiannakis, and Y.~Wu,
``Natural TTS synthesis by conditioning WaveNet on mel spectrogram
predictions,''
in \emph{Proceedings of the IEEE International Conference on Acoustics,
Speech and Signal Processing (ICASSP)},
pp.~4779--4783, IEEE, Apr. 2018.

\bibitem{li2021starganv2vc}
Y.~A.~Li, A.~Zare, and N.~Mesgarani,
``StarGANv2-VC: A diverse, unsupervised, non-parallel framework for
natural-sounding voice conversion,''
\emph{arXiv preprint arXiv:2107.10394}, 2021.

\bibitem{albadawy2020voice}
E.~A. AlBadawy and S.~Lyu,
``Voice conversion using speech-to-speech neuro-style transfer,''
in \emph{Proc. Interspeech},
2020, pp.~4726--4730.

\bibitem{mohamed2022selfsupervised}
A.~Mohamed, H.-Y.~Lee, L.~Borgholt, J.~D.~Havtorn, J.~Edin,
C.~Igel, K.~Kirchhoff, S.-W.~Li, K.~Livescu, L.~Mangu,
T.~N.~Sainath, and S.~Watanabe,
``Self-supervised speech representation learning: A review,''
\emph{IEEE Journal of Selected Topics in Signal Processing},
vol.~16, no.~6, pp.~1179--1210, 2022.

\bibitem{sen2025noiseaware}
U.~Sen, A.~Luqman, and A.~Chattopadhyay,
``Toward noise-aware audio deepfake detection: Survey, SNR-benchmarks, and practical recipes,''
\emph{arXiv preprint arXiv:2512.13744}, 2025.

\bibitem{zhang2022c3dino}
C.~Zhang and D.~Yu,
``C3-DINO: Joint contrastive and non-contrastive self-supervised learning for speaker verification,''
\emph{IEEE Journal of Selected Topics in Signal Processing},
vol.~16, no.~6, pp.~1273--1283, 2022.

\bibitem{gao2026context}
C.~Gao, M.~Postiglione, J.~Baldwin, N.~Denisenko, I.~Gortner,
L.~Fosdick, and V.~S.~Subrahmanian,
``Context and transcripts improve detection of deepfake audios of public figures,''
\emph{arXiv preprint arXiv:2601.13464}, 2026.

\bibitem{zahran2023finetuning}
A.~I. Zahran, A.~A. Fahmy, K.~T. Wassif, and H.~Bayomi,
``Fine-tuning self-supervised learning models for end-to-end pronunciation scoring,''
\emph{IEEE Access},
vol.~11, pp.~112650--112663, 2023.

\bibitem{ge2017optimal}
W.~Ge, Z.~Fazal, and E.~Jakobsson,
``Using optimal f-measure and random resampling in gene ontology enrichment calculations,''
\emph{Frontiers in Applied Mathematics and Statistics},
vol.~5, p.~20, 2019.

\bibitem{tuglus2009modified}
C.~Tuglus and M.~J. van~der~Laan,
``Modified FDR controlling procedure for multi-stage analyses,''
\emph{Statistical Applications in Genetics and Molecular Biology},
vol.~8, no.~1, Art.~12, 2009.

\bibitem{hanilci2026cyclostationarity}
C.~Hanil{\c{c}}i, M.~Sahidullah, and T.~Kinnunen,
``Cyclostationarity analysis as a complement to self-supervised representations for speech deepfake detection,''
\emph{arXiv preprint arXiv:2603.03921}, 2026.

\bibitem{muller2026trainingfree}
B.~Muller, A.~A. Ortiz~Barra{\~n}{\'o}n, and L.~Roberts,
``Training-free cross-lingual dysarthria severity assessment via phonological subspace analysis in self-supervised speech representations,''
\emph{medRxiv}, 2026.

\bibitem{cowling1996bootstrap}
A.~Cowling, P.~Hall, and M.~J.~Phillips,
``Bootstrap confidence regions for the intensity of a Poisson point
process,''
\emph{Journal of the American Statistical Association},
vol.~91, no.~436, pp.~1516--1524, 1996.

\bibitem{pytorch2019}
A.~Paszke, S.~Gross, F.~Massa, A.~Lerer, J.~Bradbury, G.~Chanan,
T.~Killeen, Z.~Lin, N.~Gimelshein, L.~Antiga, A.~Desmaison,
A.~K{\"o}pf, E.~Yang, Z.~DeVito, M.~Raison, A.~Tejani,
S.~Chilamkurthy, B.~Steiner, L.~Fang, J.~Bai, and S.~Chintala,
``PyTorch: An imperative style, high-performance deep learning library,''
in \emph{Advances in Neural Information Processing Systems},
vol.~32, 2019.

\bibitem{dwork2012fairness}
C.~Dwork, M.~Hardt, T.~Pitassi, O.~Reingold, and R.~Zemel,
``Fairness through awareness,''
in \emph{Proceedings of the 3rd Innovations in Theoretical Computer Science Conference},
2012, pp.~214--226.

\bibitem{nallaguntla2026phoneme}
V.~Nallaguntla, S.~Kshirsagar, and A.~R.~Avila,
``Phoneme-Level Deepfake Detection Across Emotional Conditions Using Self-Supervised Embeddings,''
\emph{arXiv preprint arXiv:2605.03079}, 2026.

\bibitem{pessach2022review}
D.~Pessach and E.~Shmueli,
``A review on fairness in machine learning,''
\emph{ACM Computing Surveys},
vol.~55, no.~3, pp.~1--44, 2022.

\bibitem{verma2018fairness}
S.~Verma and J.~Rubin,
``Fairness definitions explained,''
in \emph{Proceedings of the International Workshop on Software Fairness},
pp.~1--7, May 2018.

\bibitem{hellman2020measuring}
D.~Hellman,
``Measuring algorithmic fairness,''
\emph{Virginia Law Review},
vol.~106, no.~4, pp.~811--866, 2020.

\bibitem{wachter2021bias}
S.~Wachter, B.~Mittelstadt, and C.~Russell,
``Bias preservation in machine learning: The legality of fairness metrics
under EU non-discrimination law,''
\emph{West Virginia Law Review},
vol.~123, no.~3, pp.~735--790, 2021.

\bibitem{jain2021inclusive}
N.~Jain and H.~Wang,
``Inclusive speaker verification with adaptive thresholding,''
\emph{arXiv preprint arXiv:2111.05501}, 2021.

\bibitem{pleiss2017fairness}
G.~Pleiss, M.~Raghavan, F.~Wu, J.~Kleinberg, and K.~Q. Weinberger,
``On fairness and calibration,''
in \emph{Advances in Neural Information Processing Systems},
vol.~30, 2017.

\end{thebibliography}
\end{document}